\documentclass[useAMS,usenatbib,12pt]{mn2e}
\usepackage{graphics}
\usepackage{amsfonts}
\usepackage{amsmath}
\usepackage{multicol}
\usepackage{layout}
\usepackage{amssymb}
\usepackage{epsfig}

\usepackage{natbib}
\usepackage{aas_macros}

\DeclareGraphicsExtensions{.eps}

\newcommand{\lam}{$\lambda$}
\newcommand{\lamc}{$\lambda_c$}
\newcommand{\llamc}{$\log \lambda_c$}
\newcommand{\PL}{$P(\lambda)$}

\newcommand{\Lmine}{L_{\rm min}}

\newcommand{\mh}{$M_{\rm halo}$}
\newcommand{\mbh}{$M_{\rm BH}$}

\newcommand{\mbhe}{M_{\rm BH}}
\newcommand{\mbhex}{M_{\rm BH, MAX}}
\newcommand{\msun}{$M_{\odot}$}
\newcommand{\msune}{M_{\odot}}

\newcommand{\Nact}{$N_{\rm act}(M_{\rm BH},z)$}
\newcommand{\Nacte}{N_{\rm act}(M_{\rm BH},z)}

\newcommand{\ergs}{${\rm erg \, s^{-1}}$}
\newcommand{\ergse}{{\rm erg \, s^{-1}}}
\newcommand{\sigL}{$\Sigma_{\lambda}$}

\newcommand{\drhobh}{$d\rho_{\rm BH}/dt(z)$}
\newcommand{\rhobh}{$\rho_{\rm BH}(z)$}

\newcommand{\del}{$\delta$}
\newcommand{\GzM}{G$(z,M_{\rm BH})$}

\newcommand{\Gz}{G$(z)$}

\newcommand{\GPzM}{G+P$(z,M_{\rm BH})$}

\newcommand{\PhiBH}{$\Phi_{\rm BH}(M_{\rm BH},z)$}
\newcommand{\Phibh}{$\Phi_{\rm BH}(M_{\rm BH},z)$}
\newcommand{\Pobs}{$P_{\rm obs}(\lambda|z,L_{\rm min},L_{\rm max})$}
\newcommand{\UM}{$U(M_{\rm BH},z)$}

\newcommand{\PLMz}{$P(\lambda|M_{\rm BH},z)$}
\newcommand{\PLzM}{$P(\lambda|M_{\rm BH},z)$}
\newcommand{\epsi}{$\epsilon$}

\def\lam{$\lambda$}

\def\ls{\lower 2pt \hbox{$\;\scriptscriptstyle \buildrel<\over\sim\;$}}
\def\gs{\lower 2pt \hbox{$\;\scriptscriptstyle \buildrel>\over\sim\;$}}
\def\ergs{${\rm erg\, s^{-1}}$}

\def\ls{\lower 2pt \hbox{$\;\scriptscriptstyle \buildrel<\over\sim\;$}}
\def\gs{\lower 2pt \hbox{$\;\scriptscriptstyle \buildrel>\over\sim\;$}}
\def\mbh{$M_{\rm BH}$}

\title[Accretion-Driven models of Black Holes]{Accretion-Driven Evolution
of Black Holes: Eddington Ratios, Duty Cycles, and Active Galaxy Fractions}
\author[F. Shankar et. al.]{Francesco Shankar$^{1,2}$
\thanks{E-mail:$\;$francesco.shankar@obspm.fr, \, Marie Curie Fellow}
David H. Weinberg$^{3}$, and Jordi Miralda-Escud\'{e}$^{4,5}$\\
$^{1}$GEPI, Observatoire de Paris, CNRS, Univ. Paris Diderot, 5 Place Jules Janssen, F-92195 Meudon, France\\
$^{2}$Max-Planck-Institut f$\ddot{u}$r Astrophysik, Karl-Schwarzschild-Str. 1,
D-85748 Garching, Germany\\
$^{3}$Department of Astronomy, Ohio State University, McPherson Laboratory,
140 W. 18th Ave., Columbus, OH 43210-1173, USA\\
$^{4}$Instituci\'o Catalana de Recerca i Estudis Avan\c cats,
Barcelona, Spain\\
$^{5}$ Institut de Ci\`{e}ncies del Cosmos, Universitat de Barcelona,
Barcelona Spain}
\begin{document}
\date{}
\pagerange{\pageref{firstpage}--
\pageref{lastpage}} \pubyear{2012}
\maketitle
\label{firstpage}

\voffset=-0.5in


\begin{abstract}
We develop semi-empirical models of the supermassive black hole and active
galactic nucleus (AGN) populations, which incorporate the black hole growth implied by the observed AGN luminosity function assuming a radiative efficiency $\epsilon$ and a distribution of Eddington ratios $\lambda$. By generalizing these continuity-equation models to allow a distribution $P(\lambda|\mbhe,z)$ we
are able to draw on constraints from observationally estimated $\lambda$ distributions and active galaxy fractions while accounting for the luminosity thresholds of observational samples. We consider models with a Gaussian distribution of log$\,\lambda$ and Gaussians augmented with a power-law tail
to low $\lambda$. Within our framework, reproducing the high observed AGN fractions at low redshift requires a characteristic Eddington ratio $\lambda_c$ that declines at late times, and matching observed Eddington ratio distributions
requires a $P(\lambda)$ that broadens at low redshift. To reproduce the observed
increase of AGN fraction with black hole or galaxy mass, we also require a $\lambda_c$ that decreases with increasing black hole mass, reducing the AGN
luminosity associated with the most massive black holes. Finally, achieving a
good match to the high mass end of the local black hole mass function requires
an increased radiative efficiency at high black hole mass. We discuss the potential impact of black hole mergers or a $\lambda$-dependent bolometric
correction, and we compute evolutionary predictions for black hole
and galaxy specific accretion rates.
Despite the flexibility of our framework,
no one model provides a good fit to all the data we consider;
it is particularly difficult to reconcile the
relatively narrow $\lambda$ distributions and low duty cycles estimated for
luminous broad-line AGN with the broader $\lambda$ distributions and higher
duty cycles found in more widely selected AGN samples, which typically
have lower luminosity thresholds.
\end{abstract}




\section{Introduction}\label{sec|intro}

Supermassive black holes reside at the center of most, if not all,
massive galaxies.  The masses of black holes are tightly
correlated with properties of their galactic hosts, especially
the velocity dispersions and masses of their stellar bulges
(e.g., Ferrarese \& Merritt 2000; Gebhardt et al.\ 2000;
H\"aring \& Rix 2004).  If black holes grow principally
by radiatively efficient accretion, then the statistics
of quasars and active galactic nuclei (AGN) can be used to
track the growth of the black hole population.
One common approach to modeling this growth uses the
black hole ``continuity equation'' (Cavaliere et al.\ 1971;
Small \& Blandford 1992).  Simple versions of such models
have two free parameters: the radiative efficiency $\epsilon$,
which relates mass accretion rate to luminosity, and the Eddington
ratio $\lambda = L/L_{\rm Edd}$, which relates luminosity
to black hole mass. With the simplifying assumption that all active
black holes have the same fixed values of $\epsilon$ and $\lambda$, and
that each black hole has a duty cycle (i.e., a
probability of being active at a given time)
that depends on mass and redshift, \UM ,
one can use the observed AGN luminosity function to infer the
average rate at which mass is being added to each mass range
of the black hole population, thus evolving the black hole mass
function forward in time.  Model assumptions can be tested
by comparing to observational estimates of the local ($z=0$)
black hole mass function (e.g., Salucci et al. 1998; Marconi et al.\ 2004;
Shankar et al.\ 2004; Shankar, Weinberg \& Miralda-Escud\'e 2009, hereafter
SWM).

  The models presented in SWM made use of this simplifying assumption of
fixed values of $\epsilon$ and $\lambda$. Our key findings were
{\it (i)} that models with $\epsilon \approx 0.07$ and
$\lambda \approx 0.25$ yield a reasonable match to the
local black hole mass function, though these specific
inferred parameter values are sensitive to remaining uncertainties in the
bolometric
AGN luminosity function, and
{\it (ii)} that reproducing the observed luminosity function with these models
required ``downsizing'' evolution, with the duty
cycle for high mass black holes declining more rapidly with time than the
duty cycle for low mass black holes.
Additional constraints on duty cycles (apart from their relation to the
fraction of active galaxies) can be inferred from AGN and
quasar clustering
(Haiman \& Hui 2001; Martini \& Weinberg 2001), although these
constraints depend on the assumed level of scatter
between AGN luminosity and the mass of the host dark matter
halo.  Reproducing the strong observed clustering of
$z=4$ quasars (Shen et al.\ 2007) requires duty cycles
close to one and minimal scatter between luminosity and
halo mass (White, Martini \& Cohn 2008; Shankar et al. 2010b).
However, reconciling the duty cycles predicted by SWM with the
lower redshift clustering ($z\approx 1.5$) measured by
Shen et al.\ (2009) requires significant scatter
($\Sigma \approx 0.5\,$dex) between luminosity and halo
mass to lower the predicted clustering amplitude
(Shankar, Weinberg \& Shen 2010c).

In this paper we consider more general continuity-equation models than
those in SWM, incorporating a distribution of Eddington ratios \PL,
as well as
allowing for possible redshift evolution or black hole mass dependence
of characteristic $\lambda$ values or average radiative efficiency.
The motivation for these more general models is to make contact with
a broader range of observations that offer additional constraints on
AGN and black hole evolution.  Most obviously, many authors have
now used black hole masses inferred from linewidth measurements
or host galaxy properties to directly estimate Eddington ratios,
finding evidence for broad \PL\ distributions that change with
redshift and black hole mass (see Section~\ref{subsec|CompareWithObservedPL}
below for observational references).  Active galaxy fractions provide
another set of important constraints on models.  In single-$\lambda$
models the duty cycle \UM\ --- the fraction of black holes of mass $\mbhe$
at redshift $z$ that are active at a given time ---
follows simply from the ratio of the observed luminosity function
to the calculated black hole mass function (see SWM).
With a broad \PL, the definition of duty cycle depends on the
adopted threshold in luminosity or Eddington ratio, and because
observational samples inevitably include these thresholds, the observed
active galaxy fraction also tests models of \PL.
 Other authors have recently implemented continuity equation models
with broad Eddington ratio distributions
(Merloni \& Heinz 2008; Yu \& Lu 2008; Cao 2010),
but their studies are limited to a few specific models.
Our implementation here draws on some of the techniques introduced
by Cao (2010) and also on techniques and ideas from Steed \& Weinberg (2003),
who attempted a broad-ranging study of black hole evolution within a
general continuity equation framework.
The broad \PL\ models described here also provide a more
general framework for modeling quasar and AGN clustering,
a topic we will examine in future work.

Phenomenological models of the black hole population like
those in this paper complement models based on hydrodynamic
simulations (e.g., di Matteo et al.\ 2005) or
semi-analytic models that tie quasar activity to major
mergers of galaxies or dark matter halos
(e.g., Kauffmann \& Haehnelt 2000; Wyithe \& Loeb 2003;
Hopkins et al. 2006).  The numerical and semi-analytic
approaches adopt a more complete physical scenario for the
mechanisms that drive black hole accretion, and they tie
AGN to the underlying galaxy population by construction.
However, it can be difficult to interpret the significance
of discrepancies with observations, or to know whether observational
successes imply that the physical assumptions are correct.
Phenomenological models are free to be driven by the
data, within the constraints imposed by the adopted parameterization.

The next section describes our methods for computing black
hole evolution and duty cycles, and it summarizes the SWM
estimate of the bolometric luminosity function that is our
basic observational input.
All of our models reproduce this luminosity function by construction.
In Section~\ref{sec|broadPL} we
describe our three basic models for \PL\ --- a
$\delta$-function, a Gaussian in $\log\lambda$, and a
Gaussian plus a power-law tail to low $\lambda$ ---
then examine their impact on mass function evolution and
conduct a first comparison to observed $\lambda$ distributions.
In Section~\ref{sec|ActiveFractions} we turn our focus to
active galaxy fractions and discuss how data might favour
redshift evolution and possibly mass dependence of \PL.
In Section~\ref{sec|OtherConstraints}
we revisit the comparison to observed $\lambda$ distributions,
the impact of black hole mergers at the rate predicted by
hierarchical galaxy formation models, the relation between specific
black hole growth and specific star formation rate,
and the impact of adopting a $\lambda$-dependent bolometric correction.
In Section~\ref{sec|discu} we discuss our results highlighting some internal tensions
between data sets and key tensions between models and data.
We conclude in Section~\ref{sec|conclu}.
The appendices present some details of our calculational methods
and of our estimates of observed active galaxy fractions.
Throughout the paper we use cosmological parameters $\Omega_m=0.30$,
$\Omega_\Lambda=0.70$, and $h\equiv H_0/100\, {\rm km\, s^{-1}\, Mpc^{-1}}=0.7$,
and we compute the linear matter power spectra running the
Smith et al.\ (2003) code with $\Gamma=0.19$, and $\sigma_8=0.8$.

\section{Formalism}\label{sec|formalism}

\subsection{Evolving the black hole mass function}
\label{subsec|BHMFevol}

To study the global evolution of the black hole population through
time, we develop models in which the black hole mass function is
self-consistently evolved via the continuity equation,
\begin{equation}
\frac{\partial n_{\rm BH}}{\partial
t}(M_{\rm BH},t)=-\frac{\partial (\langle \dot{M}_{\rm BH}\rangle
n_{\rm BH}(M_{\rm BH},t))}{\partial M_{\rm BH}}\,,
    \label{eq|conteq}
\end{equation}
where $\langle \dot{M}_{\rm BH}\rangle$ is the mean accretion
rate (averaged over the active and inactive populations) of the
black holes of mass \mbh\ at time $t$
(see, e.g.,
Cavaliere et al. 1971; Small \& Blandford 1992; Yu \& Tremaine 2002; Steed
\& Weinberg 2003; Marconi et al. 2004; Merloni 2004; Yu \& Lu 2004; Tamura
et al. 2006;
SWM; Shankar 2009; Raimundo \& Fabian 2009; Kisaka \& Kojima 2010).
This evolution is equivalent
to the case in which every black hole grows constantly at the mean
accretion rate $\langle \dot{M}_{\rm BH}\rangle$. In practice, individual
black holes turn on and off, but the mass function evolution depends only on
the mean accretion rate as a function of mass. Note that Eq.~(\ref{eq|conteq})
neglects any contribution from black hole
mergers, which do not add mass to the black hole population but can alter
the mass function by redistribution.
In Section~\ref{subsec|Mergers} we discuss the impact
of including black hole mergers at the rate predicted by hierarchical models
of structure formation.

We first define the Eddington ratio,
\begin{equation}
\lambda \equiv {L \over L_{\rm Edd}},
\end{equation}
where
\begin{equation}
L_{\rm Edd}\equiv l \mbhe =1.26\times 10^{38}\left(\frac{M_{\rm BH}}{{\rm
M_{\odot}}}\right) \, {\rm erg\, s^{-1}}\,
    \label{eq|Lbol}
\end{equation}
is the Eddington (1922) luminosity computed for Thomson scattering
opacity and pure hydrogen composition. The growth rate of an active
black hole of mass \mbh\ with Eddington ratio \lam\ is $\dot{M}_{\rm
BH}=$\mbh$/t_{\rm ef}$,
where the \emph{e}-folding time is (Salpeter 1964)
\begin{equation}
t_{\rm ef}=4\times 10^8\left(\frac{f}{\lambda}\right)\, {\rm yr}
  \equiv {t_s \over \lambda}
 ~,
    \label{eq|tefold}
\end{equation}
with
\begin{equation}
f=\epsilon/(1-\epsilon) \quad \hbox{and} \quad
\epsilon=\frac{L}{\dot{M}_{\rm inflow}c^2}. \label{eq|epsilon}
\end{equation}
The radiative efficiency $\epsilon$ is conventionally defined with
respect to the large scale mass inflow rate, but the black hole
growth rate $\dot{M}_{\rm BH}$ is smaller by a factor of $(1-\epsilon)$
because of radiative losses.

We then define the probability for a black hole of a given mass
\mbh\ to accrete at the Eddington ratio \lam\,
per unit $\log\lambda$, at redshift $z$ as
\PLzM, normalized to unity, i.e.,
$\int d\log \lambda \, P(\lambda|M_{\rm BH},z)=1$.\footnote{Throughout
the paper, log denotes base-10 logarithm and ln denotes natural logarithm.}
This is the
quantity that can be related to the observed Eddington ratio distributions
discussed in Section~\ref{sec|intro}.
The average growth
rate of all black holes can be computed by
convolving the input \PLzM\ with the duty cycle \UM,
\begin{equation}
\langle \dot{M}_{\rm BH}\rangle=\int d\log \lambda \, P(\lambda|M_{\rm
BH},z)\lambda \, U(M_{\rm BH},z)\,  \frac{M_{\rm BH}}{t_{\rm s}}\, ,
\label{eq|MdotAve}
\end{equation}
where the integral extends over all allowed values of $\lambda$.
A physically consistent model must have $U(\lambda,M_{\rm BH},z)\le 1$
for all \mbh\ and $z$.
In models where the Eddington ratio has a single value $\lambda_1$,
$P(\lambda|M_{\rm BH},z)=\lambda_1 \delta(\lambda - \lambda_1)$,
the duty cycle is simply the ratio of the luminosity and black hole mass
functions,\footnote{Following SWM, we will always use the symbol $\Phi(x)$
to denote mass and luminosity
functions in logarithmic units of $L$ or \mbh,
i.e. $\Phi(x)d\log x=n(x)x \ln(10)dx$,
where $n(x)dx$ is the comoving space density of black holes in the
mass or luminosity range $x\rightarrow x+dx$, in units of ${\rm Mpc^{-3}}$
for $h=0.7$.
Unless otherwise stated, all masses are in units of $M_\odot$ and all
luminosities are bolometric
and in units of \ergs; e.g., $\log L > 45$ refers to quasars with bolometric
luminosity $L > 10^{45} \ergse$.}
\begin{equation}
U(M_{\rm BH},z)=\frac{\Phi(L,z)}{\Phi_{\rm BH}(M_{\rm BH},z)}\, ,
\qquad\qquad L=\lambda l M_{\rm BH}\, .
    \label{eq|P0general}
\end{equation}

\begin{figure*}
    \includegraphics[width=15truecm]{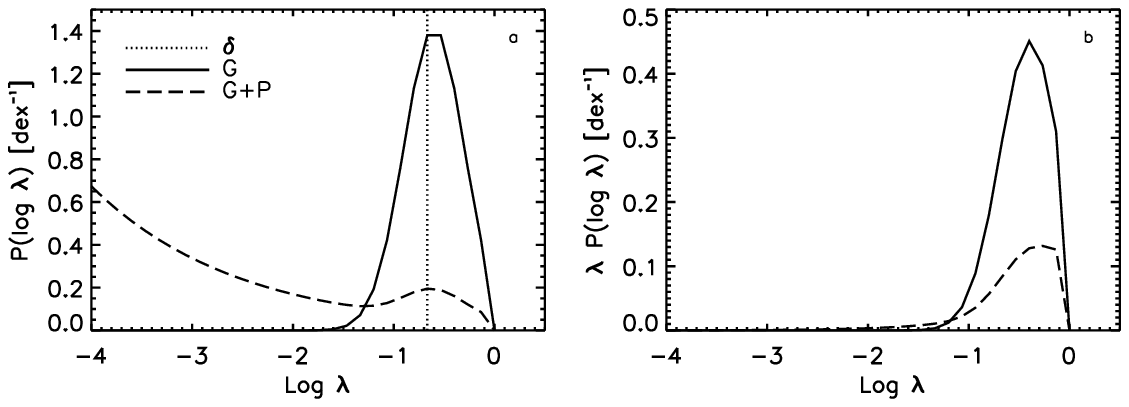}
    \caption{\emph{Left}: Comparison among the three \PL\ distributions
    taken as a reference throughout the paper.
The \emph{dotted} line is the \del-model centered on \llamc$=-0.6$, the \emph{solid} line is
the G-model, a Gaussian with dispersion of \sigL$=0.3$ dex and centered around
the same value of \llamc$=-0.6$, and
the \emph{long-dashed} line is the G+P-model, characterized by the same Gaussian plus
a power-law with slope $\alpha=-0.3$ normalized
to have the same value as the Gaussian at $\log \lambda=\log \lambda_c-
\log \Sigma_{\lambda}$. Note that
we define our distributions in $\log \lambda$ but refer to them in
figures and text as $P(\lambda)$,
and that all distributions are truncated at $\lambda=1$. \emph{Right}:
Dimensionless emissivity per $\log\lambda$ interval,
$\lambda P(\lambda)$, for the G and G+P distributions.
Note that even the G+P model has its emissivity peak near
the peak of the Gaussian component.}
\label{fig|comparePLambda}
\end{figure*}

  This paper presents predictions from several models that take as
input the observed luminosity function $\Phi(L,z)$, an assumed
Eddington ratio distribution of active black holes, \PLzM, and an
assumed radiative efficiency $\epsilon$, to solve
Eq.~(\ref{eq|conteq}) forwards in time. At each timestep,
we use the black hole mass function $\Phi_{\rm BH}(M_{\rm BH},z)$ derived
from the previous timestep (or from an assumed initial
condition in the first timestep), and
we compute the duty cycle \UM\ by numerically solving the equation
\begin{equation}
\Phi(L,z)=\int d\log \lambda P(\lambda|M_{\rm BH},z) U(M_{\rm BH},z)
\Phi_{\rm BH}(M_{\rm BH},z)\,
\label{eq|PhiLLambda}
\end{equation}
with $\mbhe = L/(\lambda l)$.
We then compute the average growth rate as a function of black hole mass via
Eq.~(\ref{eq|MdotAve}) and, finally,
update the black hole masses and black hole mass function via the
continuity equation (Eq.~\ref{eq|conteq}).

  Details of our numerical implementation are given in
Appendix~\ref{app|solvingContEq}.
Directly solving Eq.~(\ref{eq|PhiLLambda}) for $U(M_{\rm
BH},z)$ is feasible for a discrete distribution of Eddington ratios.
However, for all of the models discussed
in this work, we will adopt \emph{continuous} \PLzM\ distributions,
which can be more efficiently handled following the method
described by Steed \& Weinberg (2003) and Cao (2010).
The latter is based on solving
Eq.~(\ref{eq|PhiLLambda}) for the full mass function of \emph{active}
black holes
\begin{equation}
\Nacte=\Phi_{\rm BH}(M_{\rm BH},z) \times U(M_{\rm BH},z) \, .
\label{eq|NactDef}
\end{equation}
More specifically,
at any redshift $z$ a parameterized active mass function \Nact\ is derived by
directly fitting\footnote{As discussed in Appendix~\ref{app|solvingContEq},
we adopt a double power-law form for the input \Nact, which is a good
(though not perfect) approximation of the functional form adopted for the
AGN luminosity function
(see Section~\ref{subsec|AGNLF}).
Our procedure allows us to reproduce the AGN
luminosity function at the $\sim 5\%$ level,
adequate precision given the statistical uncertainties
in the data.}
the AGN luminosity function $\Phi(L,z)$ at the same redshift via
Eq.~(\ref{eq|PhiLLambda}). The function
\Nact\ is then inserted in Eq.~(\ref{eq|conteq}) by replacing
$\langle \dot{M}_{\rm BH}\rangle
n_{\rm BH}(M_{\rm BH},t)=\int d\log \lambda P(\lambda|M_{\rm BH},z)
\mbhe \Nacte/t_{\rm s}$,
thus allowing the computation of the full black hole mass function \PhiBH\ and
the duty cycle $U(M_{\rm BH},z)=\Nacte/\Phi_{\rm BH}(M_{\rm BH},z)$.

Eq.~(\ref{eq|conteq}) is solved by imposing an initial condition
that we fix, as in SWM, by assuming a constant value of
the initial duty cycle.
We usually choose this to be 0.5 at $z=6$ for all masses,
but in some models we lower it to
0.1 to allow the duty cycles at later times
to be always lower than unity.
The results at later redshifts are insensitive to the initial
conditions for mass bins that have experienced substantial accretion
growth (Marconi et al.\ 2004; SWM).
Note that Eq.~(\ref{eq|conteq}) could be solved backwards
in time as well by forcing the initial condition to be the
local black hole mass function (e.g., Merloni 2004).
In this case, incorrect assumptions about the physical parameters
or present day mass function manifest themselves as unphysical
mass functions at earlier redshifts (e.g., negative space densities).
Given the still substantial uncertainties in the local mass function
(e.g., SWM; Vika et al.\ 2009; Shankar et al., in prep.) and our
desire to explore a wide range of physical models, we prefer to
evolve forward in time and compare to the local mass function
as one of several observational constraints.

\subsection{Input AGN Luminosity Function}\label{subsec|AGNLF}

A full discussion of our adopted AGN luminosity function appears
in SWM; here we provide a brief summary.\footnote{The
full tables of the input
AGN luminosity functions, and also black hole mass functions and duty cycles (for a set of
representative models), can be found
in electronic format at {\tt http://mygepi.obspm.fr/$\sim$fshankar/}.}
We adopt the Ueda et al. (2003) fit to
the number of sources per unit volume per dex of luminosity $\log
L_X=\log L_{2-10\, {\rm keV}}$, composed of a
double power-law and an evolution term:
\begin{equation}
\Phi(L_X,z) = e(z,L_X)\, \frac{A}{
 \left( \frac{L_X}{L^*}\right)^{\gamma_1} +
 \left( \frac{L_X}{L^*}\right)^{\gamma_2} } \, ,
    \label{eq|U03}
\end{equation}
where
\begin{equation}
e(L_X,z)=\left\{
  \begin{array}{ll}
    (1+z)^{p_1}                      & \hbox{if $z<z_c(L_X)$} \\
    (1+z_c)^{p_1}[\frac{1+z}{1+z_c(L_X)}]^{p_2} & \hbox{if $z\ge z_c(L_X)$\,
    ,}
  \end{array}
\right.
    \label{eq|eLz}
\end{equation}
with
\begin{equation}
z_c(L_X)=\left\{
  \begin{array}{ll}
    z_c^*                      & \hbox{if $L_X\ge L_a$} \\
    z_c^*(L_X/L_a)^{0.335}    & \hbox{if $L_X< L_a$\, .}
  \end{array}
\right.
    \label{eq|zc}
\end{equation}
The values of the parameters have been tuned
to correctly match the full set of data presented by SWM (see their Table~1).

The X-ray luminosity function of Eq.~(\ref{eq|U03}) is converted into a bolometric
luminosity function using the fit to the bolometric correction given by
Marconi et
al. (2004), $\log L/L_X=1.54+0.24\zeta+0.012\zeta^2-0.0015\zeta^3$,
with $\zeta=\log L/L_{\odot}-12$, and $L_{\odot}=4\times 10^{33}\
{\rm erg\, s^{-1}}$.  We have changed
our procedure slightly with respect to SWM, dropping the convolution
with 0.2-dex Gaussian scatter in bolometric correction, because the
assumption that this scatter is uncorrelated with X-ray luminosity
seems insecure.  However, this change makes minimal difference to
the derived luminosity function, and our results using the original
SWM fit would not be noticeably different.

The luminosity function data that we fit also include
a number density of Compton-thick sources in each luminosity bin equal to the number of sources in the column
density range $23\le \log N_H/{\rm cm^{-2}}\le 24$, computed following
the Ueda et al. (2003) prescriptions for the $P(N_H|L)$ column density
distributions. We checked that our estimates of the Compton-thick number densities at high redshifts
are fully consistent with the recent results by Alexander et al. (2011) and Fiore et al. (2011).
As extensively discussed by SWM, the integrated intensity obtained
from our model AGN luminosity function is also consistent with all the available
data on the cosmic X-ray background for energies above 1 keV.

Our qualitative conclusions would not change if we adopted
the Hopkins et al. (2007) luminosity function in place of SWM,
though best-fit parameter values would be somewhat different.

\section{Effects of a broad \PL}\label{sec|broadPL}

In this section and the ones that follow, we will compare predictions
of our models to a variety of observational constraints.  If the observational
uncertainties were described by well defined statistical errors, then
the natural approach would be to determine best-fit model parameters
via $\chi^2$ minimization and compare different models via
$\Delta\chi^2$.  We have taken this approach in some of our previous
studies when we were focusing on a specific observational constraint
and a limited class of models.  However, the uncertainties in the
observational constraints we consider --- AGN luminosity functions,
the local black hole mass function, distributions of Eddington ratios,
and duty cycles estimated from active galaxy fractions --- are in all
cases dominated by systematic errors, and in some cases even the
rough magnitude of these systematics is difficult to estimate.
We discuss these issues in the text and Appendices below, and the papers cited
for each observable often discuss systematic uncertainties at length.

Given this situation, we adopt a philosophy of ``qualitative comparison
to quantitative data''.  We infer model parameters based on an overall
match to one or more sets of observational constraints, but we do not
attempt formal $\chi^2$-minimization because the errors themselves
are not well defined enough to do so.  We note where models fit observations
within a plausible range of systematic uncertainties, and where
they do not.  Our objective is to delineate the global characteristics
that successful black hole accretion models must possess to reproduce
the range of observables now emerging from deep, large area surveys of
the AGN and galaxy populations.

\subsection{Input Eddington ratio distributions}\label{subsec|PLmodels}

\begin{figure*}
    \includegraphics[width=15truecm]{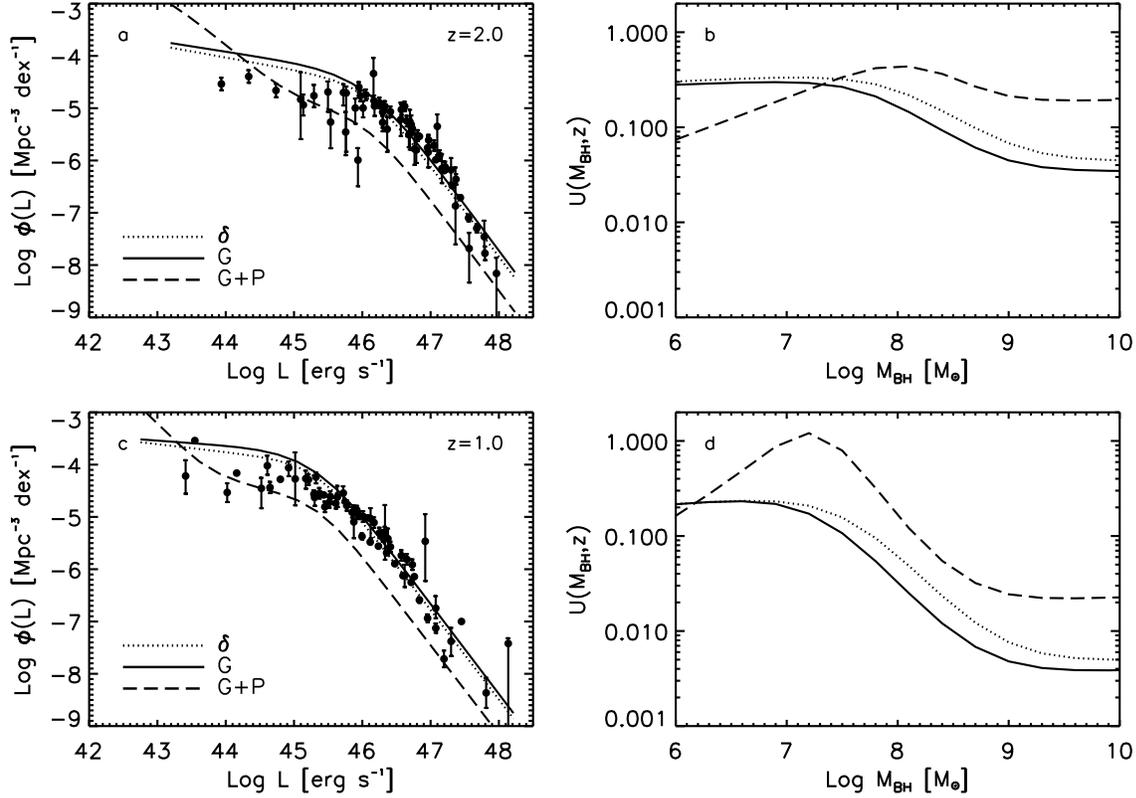}
    \caption{\emph{Left panels}: Predicted luminosity function at $z=2$
    (top) and at $z=1$ (bottom)
    for the three \PL\ distributions of Fig.~\ref{fig|comparePLambda},
    as labelled,
    when the {\it same}
    black hole mass function \PhiBH\ and duty cycle \UM\ are used, in this
    case the
    one predicted by the \del-model.
    Points show the data compilation of SWM.
    \emph{Right panels}: predicted duty cycle \UM\ at the same redshifts
    $z=2$ and $z=1$
    when, instead, the
    AGN luminosity function is \emph{fixed} to the one in SWM,
    as in our self-consistent models.
    Allowing a large fraction of sub-Eddington accretion rates,
    as in the G+P model, increases the probability and thus the duty cycle
    for more massive black holes to be active
    at different luminosities.}
    \label{fig|predictedLFandUmbh}
\end{figure*}
\begin{figure*}
    \includegraphics[width=17truecm]{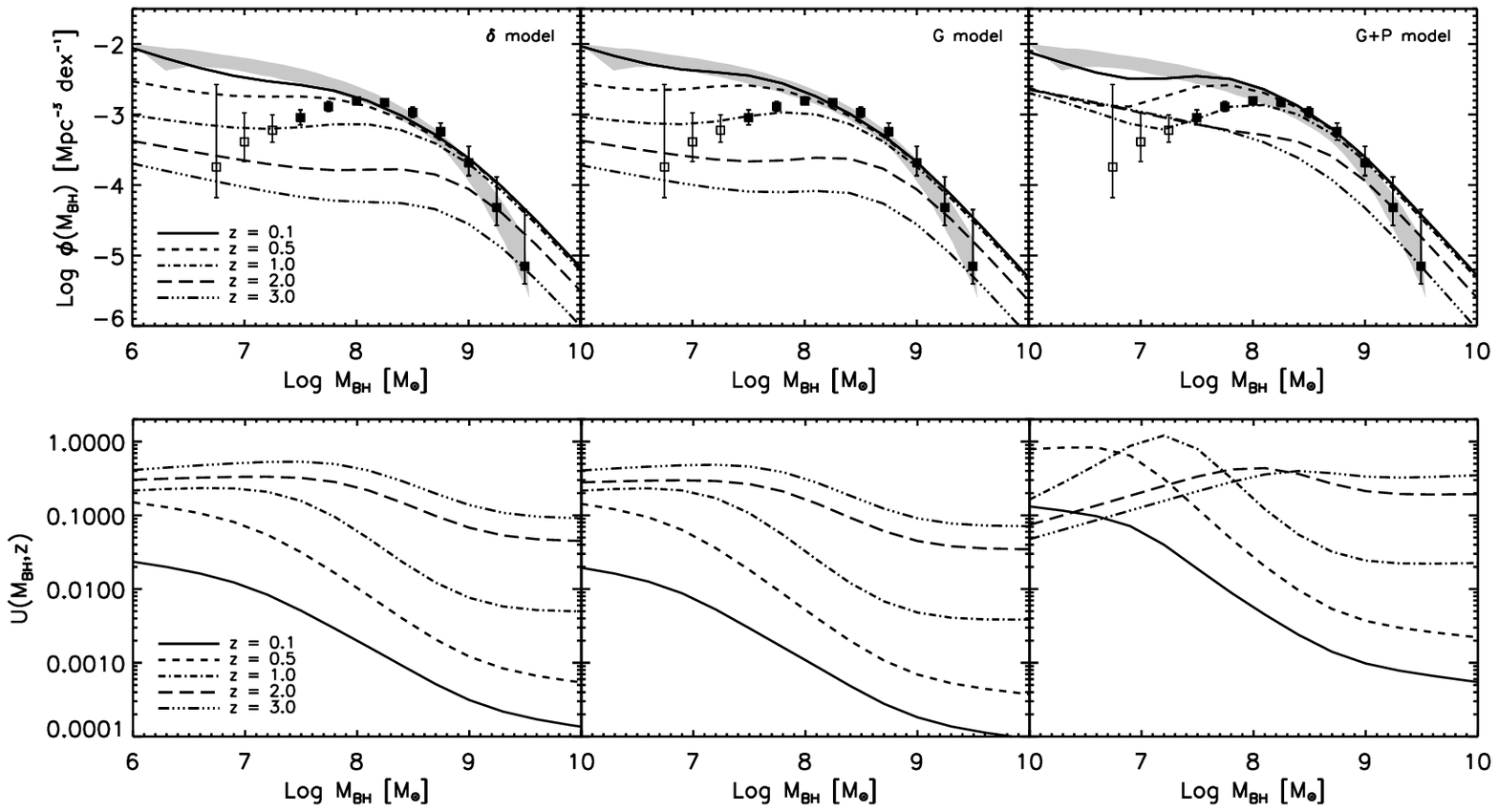}
    \caption{The \emph{left}, \emph{middle}, and \emph{right} columns refer
    to outputs
    of the \del, G, and G+P models, respectively, whose \PL\ distributions
    are shown in Fig.~\ref{fig|comparePLambda}.
    \emph{Upper panels}: Predicted black hole mass function at $z\sim 0$
    (\emph{solid} lines) compared to the local estimates by SWM (\emph{grey
    area}),
    and Vika et al. (2009; \emph{squares} with error bars,
    with the \emph{open} symbols indicating the estimates derived from
    galaxies below their reliability limit).  Other lines
    show the predicted black hole mass function at earlier redshifts, as
    labelled.  \emph{Bottom panels}: Corresponding duty cycles (for all
    sources accreting above any $\lambda>0$)
    as a function of black hole mass at different redshifts, as labelled. For
    all models a radiative efficiency of $\epsilon=0.06$ has been assumed.}
    \label{fig|BHMFandDuty}
\end{figure*}

In SWM we discussed accretion models for black holes with a single value
of \lam , examining the
effects of redshift and mass dependence of this value of \lam\
and describing the impact of uncertainties
in the input AGN luminosity function related to its observational
determination, obscured fractions, and bolometric corrections.
From the match to the local black hole mass function, we were able to set
constraints on the average input radiative efficiency and evolution in the
characteristic \lam.
In this section we study the impact of broadening the input
\PL\ distribution. To this purpose we adopt and compare the outputs
from three different distributions (shown in Fig.~\ref{fig|comparePLambda}):
\begin{itemize}
  \item the \textbf{\del-model}, centered on \llamc$=-0.6$ (dot-dashed line
  in panel \emph{a}; we choose this particular value because it provides
  a good match to the local black hole mass function);
  \item the \textbf{G-model}, a Gaussian in $\log\lambda$
  with dispersion of \sigL$=0.3$
  dex and centered around the same value of \llamc$=-0.6$ (long-dashed line
  in panel \emph{a});
  \item the \textbf{G+P-model}, characterized by the same Gaussian plus a
  power-law with slope $\alpha=-0.3$ normalized
to have the same value as the Gaussian at $\log \lambda=\log \lambda_c-
\log \Sigma_{\lambda}$ (panel \emph{b}).
\end{itemize}
The G+P distribution has a shape close to the one
inferred by Kauffmann \& Heckman (2009; see also Aird et al. 2011) in the local Universe from
the Sloan Digital Sky Survey (SDSS),
and it is also consistent with some theoretical and semi-empirical
expectations (e.g., Merloni \& Heinz 2008; Yu \& Lu 2008; Hopkins \&
Hernquist 2009; Shen 2009).
In particular, the power-law component could represent the effect of either
a steady decline in the accretion rate after a near-Eddington growth phase or
a second mode of AGN fueling triggered by secular instabilities instead
of major mergers.
As discussed by SWM and demonstrated further below, the chosen value of \llamc\ produces reasonable agreement between the local and accreted mass function, especially around $(1-3)\times 10^8\, $\msun\ where the former is best determined. Our chosen value of $\alpha$, on the other hand, was mainly derived a posteriori after some trial and error. Higher values of $\alpha$ can induce unphysical duty cycles \UM\ $>1$
in some mass bins during the evolution,
while lower values give results that are not much different from the G-model alone.
For all models we adopt a radiative efficiency of $\epsilon=0.06$, unless otherwise stated, within the range of values favoured by SWM.
In future sections and figures we will consider many variants on
these three basic models.
For reference, we provide a full list of models and their basic
properties in Table~\ref{table|models}.  We summarize the comparison
between our six primary models --- the three introduced here, plus versions
that introduce redshift and mass dependence of \PL\ ---
and observational data in Table~\ref{table|models2} (Section~\ref{sec|discu}).

\begin{table*}
\begin{tabular}{|l|l|r|}
  \hline
  Reference Models & shape \& properties & Section/Equations \\
  \hline
  \hline
  $\delta$ & $\delta$-function & Sec.~\ref{subsec|PLmodels} \\
  G & Gaussian & Sec.~\ref{subsec|PLmodels} \\
  G+P & Gaussian+Power-Law & Sec.~\ref{subsec|PLmodels} \\
  G(z) & Gaussian+$\lambda_c(z)$ & Eq.~\ref{eq|PLz} \\
  G(z,$\mbhe$) & Gaussian+$\lambda_c(z,\mbhe)$+low-$z$
  broadening+mass-dependent $\epsilon$ & Eqs.~\ref{eq|PLz}, \ref{eq|PLmassDependent}, \ref{eq|Sigma}, \ref{eq|radefficiencyMass} \\
  G+P(z,$\mbhe$) & Gaussian+$\lambda_c(z,\mbhe)$+$z$-dependent
  Power-Law+mass-dependent $\epsilon$ & Eqs.~\ref{eq|PLz}, \ref{eq|PLmassDependent}, \ref{eq|LambdaMin}, \ref{eq|radefficiencyMass}\\
  \hline
  \hline
  Test Models & shape \& properties & Ref. Eqs. \\
  \hline
  \hline
  G(z,$\mbhe$)+constant $\epsilon$ &  & \\
  G(z,$\mbhe$)+$\epsilon(z)$ & G(z,$\mbhe$)+z-dependent $\epsilon$ & Eq.~\ref{eq|radEfficiencyReds}\\
  G(z,$\mbhe$)+$K(\lambda)$ &
  G(z,$\mbhe$)+$\lambda$-dependent bolometric correction
  & Eq.~\ref{eq|BClambda} \\
  G(z,$\mbhe$)+P($z<0.7$) & G(z,$\mbhe$)+Power-Law only at $z<0.7$ & \\
  \hline
  \end{tabular}
  \caption{List of models explored in this work along with basic explanation and Sections and Equations in which they are first introduced. The models are divided into
  two groups, the ``Reference Models'', i.e., the three ones introduced and discussed in Section~\ref{subsec|PLmodels}, and the ``Test Models'' which are simply variants of the Reference Models and progressively introduced in the rest of the paper.
  The Gaussian always has
  dispersion of $\log \Sigma_{\lambda}=0.3$ dex except for the models with
  broadening, for which
  $\log \Sigma_{\lambda}$ steadily increase with decreasing redshift as
  in Eq.~(\ref{eq|Sigma}).
  The redshift and mass
  dependence in the characteristic Eddington ratio $\lambda_c$ are defined
  in Eqs.~(\ref{eq|PLz}) and (\ref{eq|PLmassDependent}). If no evolution
  is assumed then $\log \lambda_c=-0.6$. The power-law component of the
  Eddington distribution is always normalized
  to have the same value as the Gaussian at $\log \lambda=\log \lambda_c-
  \log \Sigma_{\lambda}$, and a slope in $\log \lambda$ of $\alpha=-0.3$.
  In the last quoted model the power-law has slope $\alpha=-0.9$ and is normalized
  to have the same value as the Gaussian at $\log \lambda=\log \lambda_c-
  0.2\log \Sigma_{\lambda}$.
  The input radiative efficiency can be constant or vary with mass or redshift
  (see Eqs.~\ref{eq|radefficiencyMass} and~\ref{eq|radEfficiencyReds})
  as detailed for each model.}
  \label{table|models}
\end{table*}

The left panels of Fig.~\ref{fig|predictedLFandUmbh} illustrate the effect
of varying the input Eddington ratio distribution \PL\ at fixed duty
cycle \UM\
and fixed black hole mass function \PhiBH, at $z=2$ and $z=1$.
The luminosity function
in each case is computed via the convolution of the respective \PL\
distributions
with the product of the duty cycle and black hole mass function
(Eq.~\ref{eq|PhiLLambda}). In the
case of the \del-model, the match to the input AGN luminosity
function (Section~\ref{subsec|AGNLF}) is exact.
For these panels we \emph{fix} \UM\ and \PhiBH\ to those predicted by the \del-model
(evolved in accord with the observed luminosity function),
so that we can isolate the impact of varying the input \PL\ distribution.
Switching from a $\delta$-function \PL\ to a Gaussian of
$\Sigma_\lambda=0.3\,$dex has a
mild impact on the predicted luminosity function.
However, adopting the G+P distribution at fixed \UM\
drastically lowers the luminosity function at high $L$, since the
probability is dominated by low Eddington ratios.  The shape of
$\Phi(L,z)$ parallels that of the other models above the break
luminosity, but the G+P model has a steeper faint-end slope
reflecting the contribution of low-$\lambda$ activity.

  The procedure we follow in the rest of this paper is to compute \UM\
for a given input \PL\ distribution in a way to reproduce the observed AGN
luminosity function.
The right panels of Fig.~\ref{fig|predictedLFandUmbh} show, at the same
redshifts $z=2$ and $z=1$, the duty cycles inferred by matching the SWM
luminosity function when adopting the three \PL\ distributions
and the same underlying \PhiBH, again that of the \del-model.
The duty cycles \UM\ for the \del\ and G models are similar,
but they are several times larger for the G+P model
at high masses.
As described in SWM, the duty cycles must decrease with mass in order to
reproduce the ``downsizing'' effect in AGN evolution (the characteristic
AGN luminosity increases with redshift). In the G+P model the duty
cycle can reach values close or around unity at $z \lesssim 2$ and masses $\sim 10^7 \msune$, although with low values of $\lambda$ for the majority of them. The decline of the duty
cycle in this model at masses lower than
$\mbhe \la L_*/(l\lambda_c)$, where $L_*$ is the break of the luminosity
function (Eq.~[\ref{eq|U03}]), is induced by the presence of
many AGNs radiating at low $\lambda$ from high mass black holes, which can
already account for the low-luminosity AGNs.
However, the detailed form of this decline is sensitive to the precise
shape of the AGN luminosity function and our assumed \PL.

\begin{figure*}
    \includegraphics[width=17truecm]{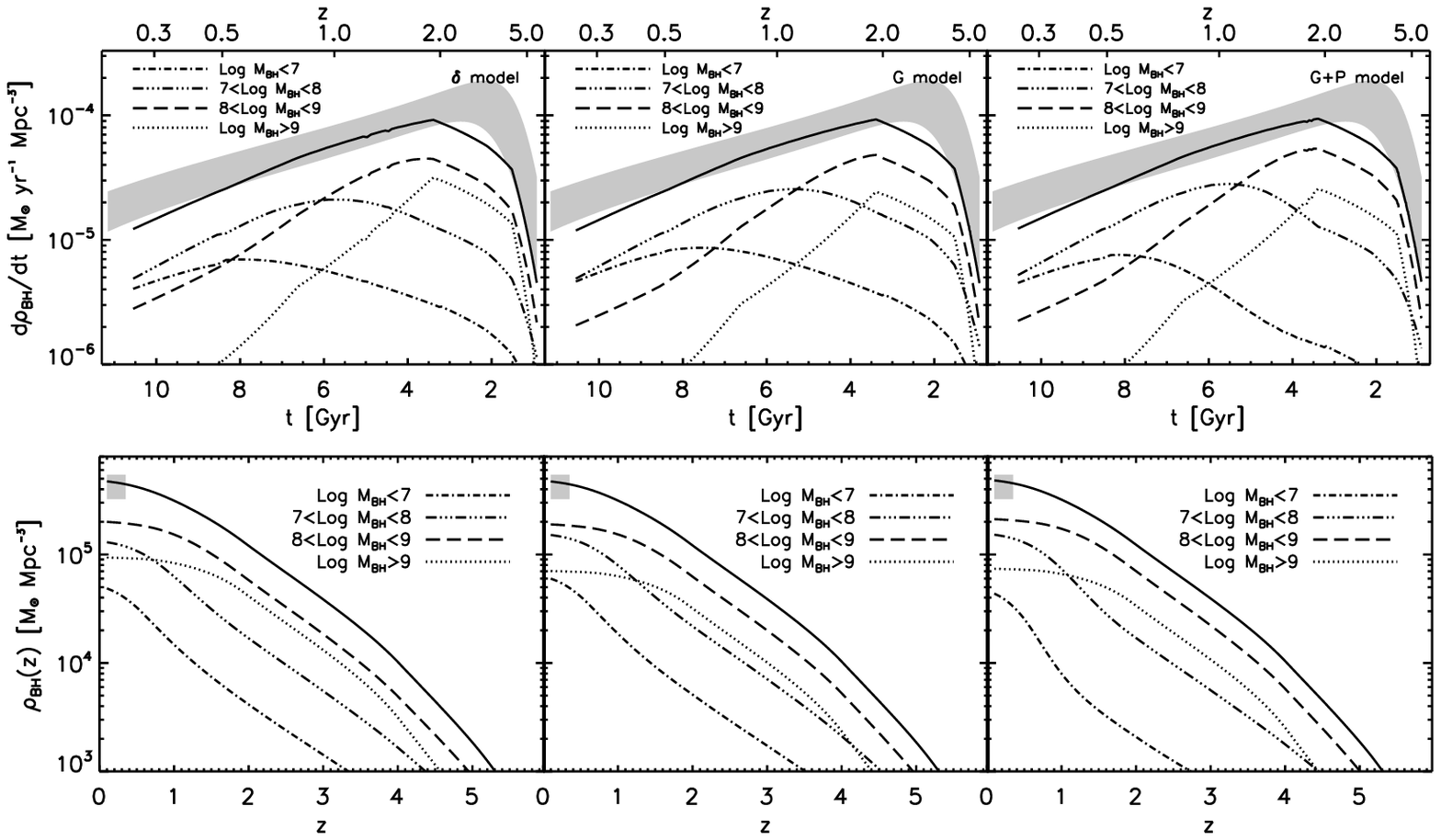}
    \caption{\emph{Upper panels}: Growth rate of the integrated black hole
    mass density as a function of time (\emph{solid} lines) and relative
    contributions of black holes of different final mass, as labelled;
    the \emph{grey area} marks the 3$\sigma$ uncertainty region of the
    cosmological star formation rate as inferred by Hopkins \& Beacom (2006),
    scaled by a factor of $6.5\times 10^{-4}$. \emph{Lower panels}: predicted
    cumulative black hole mass density as a function of redshift (\emph{solid}
    lines), and contributions of black holes of different mass, as labelled.
    The \emph{grey bar} indicates the values and systematic
	uncertainties in the total local mass density in black holes estimated
	by SWM}.
    \label{fig|RhoBHz}
\end{figure*}

\subsection{Global Accretion Histories}\label{subsec|AccretionHistories}

Fig.~\ref{fig|BHMFandDuty} shows the
evolution of the black hole mass functions and duty cycles
for the \del, G, and G+P models, respectively.
In the upper panels, grey bands show the estimate of the local black
hole mass function by SWM.  The width of this band already encompasses
a number of systematic uncertainties, but, as SWM discuss, the
inferred mass function for $\mbhe < 10^8 M_\odot$ is sensitive
to uncertainties in the treatment of spiral galaxy bulges.
Vika et al.\ (2009) have tried to address this problem by
estimating the local black hole mass function on an object by object
basis, i.e., by computing the
bulge fraction for each galaxy in a large sample, assigning
black hole masses from an \mbh-$L$ relation, and then computing
the black hole mass function applying the $V/V_{\rm max}$ method.
Their result, shown with open and filled squares in
Fig.~\ref{fig|BHMFandDuty}, agrees well with SWM at high masses,
while it turns over at $M < 10^8 M_\odot$ rather than continuing to rise
to lower masses.
We thus consider Vika et al.'s and SWM's results to broadly bracket
the still remaining uncertainties in the determination
of the local black hole mass function, and we will use both as a reference
when comparing with models.

Moving from the \del-model to the G-model makes almost no difference
to the evolution of the black hole mass function or to its $z=0$
value.  Both predictions are insensitive to our assumed initial
conditions at $z=6$ because even by $z=4$ the accumulated mass from
accretion greatly exceeds that in the initial seed population.
The G+P model differs in this regard because reproducing the $z=6$
luminosity function even with a duty cycle of 0.1 requires a high
space density of seed black holes, since many black holes are
active at low $\lambda$ values that do not contribute to the observed
range of the luminosity function.  Once the evolved black hole mass
function substantially exceeds the seed population, evolution is
similar to that of the \del\ and G models.

The clustering of quasars is a diagnostic for the space density
of the underlying black hole population, as more numerous black holes
must reside in lower mass halos that are less strongly clustered
(Haiman \& Hui 2001; Martini \& Weinberg 2001).  Applying this idea
to the strong clustering measured for $z\approx 0.4$ quasars by
Shen et al.\ (2007) in the SDSS, White et al.\ (2008) and
Shankar et al.\ (2010b) find that duty cycles close to unity
are required assuming that these highly luminous quasars have
$\lambda \ga 0.1$.  This finding is at odds with the high black hole density
required for the G+P model at high redshift.  To quantify this
statement, we have computed the large scale bias of the G+P model
using a formalism that we will describe more fully in a future paper.
In brief, we match black holes to halos and subhalos
via cumulative number matching
(similar to Conroy \& White 2012), including 0.3-dex scatter in black
hole mass about the mean relation
(see Appendix~\ref{Appendix|MbhMhaloRelation}), then assign each
black hole a luminosity based on the duty cycle and the Eddington
ratio distribution \PLMz.  We can then compute the mean large
scale halo bias for black holes in a luminosity range.  For the G+P
model, which has high black hole space density and large scatter
between luminosity and halo mass, we predict a bias of 7.5 for quasars
with $\log L > 47$ at $z=4$, which is more than $2\sigma$ discrepant
with the reference value of $12.96 \pm 2.09$ given by
Shen et al.\ (2009), though it is marginally consistent with
their lowest estimate.
We conclude that broad \PL\ distributions at very high redshifts
are observationally disfavoured, a conclusion in qualitative
agreement with the findings of Cao (2010).  If \PL\ does have a
long tail to low $\lambda$ values, it must develop after the
earliest stages of black hole growth;
we will consider models with growing power-law tails
in Section~\ref{sec|ActiveFractions}.

Duty cycles \UM\ tend to decline with redshift and with mass, for the
reasons already explained in Fig.~2. We note again the higher values
of the duty cycle in the G+P model at all redshifts. The high-redshift
curves are still affected by the initial conditions in the black hole
population at low masses.

\begin{figure*}
    \includegraphics[width=12truecm]{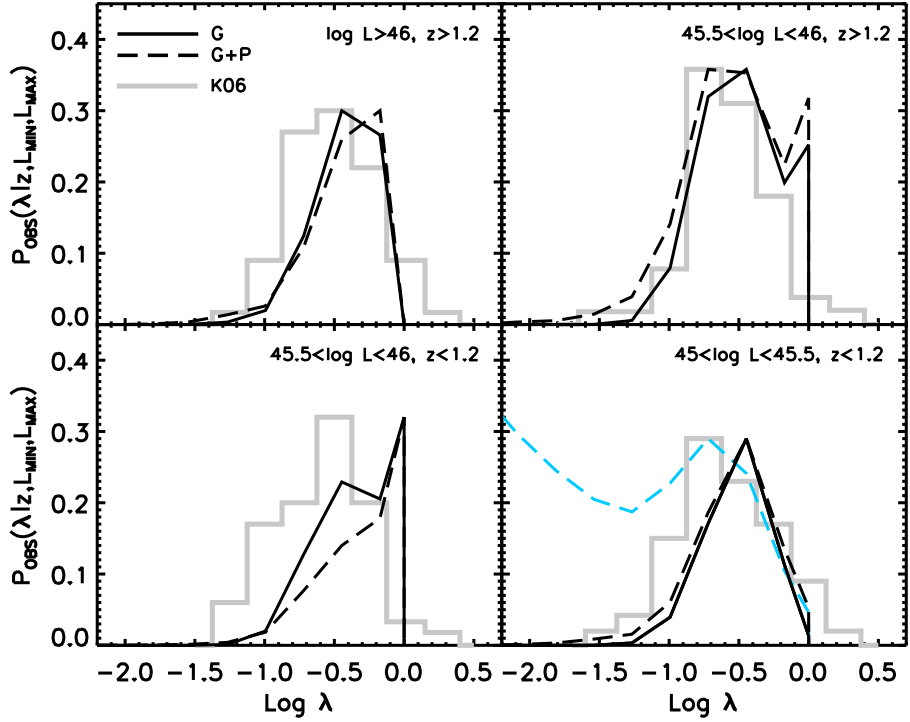}
    \caption{Eddington ratio distributions predicted by the G
    (\emph{solid} lines) and G+P (\emph{long-dashed} lines) models, compared
    to the K06 data (\emph{grey histograms}) for black
    holes in the range $10^7<\mbhe/\msune<10^{10}$ at different redshift
    and luminosity bins, as labelled. The predictions have been estimated
    at the average redshifts of $z=1$ and $z=2$ for the low and high-$z$
    subsamples, respectively, although the predictions do not strongly depend
    on these choices if no redshift dependence in the \PL\ is included. The
    \emph{cyan long-dashed line} in the lower-right panel is the G+P model
    prediction when no lower luminosity threshold is considered.}
    \label{fig|CompareKollmeier1}
\end{figure*}

Fig.~\ref{fig|RhoBHz} tracks the overall growth of the black hole
population in bins of mass.  In the upper panels, solid lines
show the growth rate of the integrated
black hole mass density \drhobh\
as a function of redshift, while the other lines show the mass
density accreted in selected bins
of \emph{current} black hole mass, as labelled. As already noted by several
groups (e.g., Marconi et al. 2004; Merloni et al. 2004; Merloni \& Heinz
2008; SWM),
the total \rhobh\ closely matches the shape of the cosmological star formation
rate (SFR), here taken from Hopkins \& Beacom
(2006; with the gray area marking their 3-$\sigma$ contours)
and re-scaled by an \emph{ad hoc} factor of $6.5\times 10^{-4}$, close
to the ratio between black hole mass and stellar mass measured in the local
Universe (e.g., H\"{a}ring \& Rix 2004).
All three models again show a clear signature of downsizing in their
accretion histories.
The accretion onto the very massive
black holes with $\log M_{\rm BH}/M_{\odot}>8$ always peaks at $z\sim 2$,
concurrent with the peak in the emissivity of luminous optical quasars
(e.g., Osmer 1982; Richards et al. 2006; Croom et al. 2009). The less
massive black
holes with $7<\log M_{\rm BH}/M_{\odot} < 8$ are characterized by a much
broader peak centered at $z\sim 1-1.5$.
The lower panels show the cumulative mass density
of all black holes with mass $10^6<\mbhe/\msune<10^{10}$ (solid lines),
and the mass density accreted onto black holes of different current mass,
as labelled.
By construction, all models share the same radiative efficiency and therefore
accumulate the same total mass densities at any time. They
differ only slightly
with respect to the total mass accumulated in different mass
bins. The solid grey square indicates the systematic uncertainties
in the total local mass density in black holes estimated by SWM.

Figures~\ref{fig|BHMFandDuty} and~\ref{fig|RhoBHz} show that the
evolution of the black hole mass function is insensitive to the
shape of \PL, provided the characteristic value $\lambda_c$ and
the input AGN luminosity function are held fixed.
This reassuring result
indicates that earlier studies assuming a single $\lambda$ value
(e.g., Marconi et al.\ 2004; Shankar et al.\ 2004; SWM) reached
robust results for mass function evolution.
This is essentially because in our G+P model,
with power-law slope $\alpha=-0.3$,
the emissivity per logarithmic
interval of $\lambda$ is dominated by the Gaussian peak
(Fig.~\ref{fig|comparePLambda}b).
A steeper slope would yield duty cycles greater than unity and is
therefore not allowed.
Even our G+P model is disfavored at high redshifts because
the observed quasar clustering implies that the most massive black
holes should be active at large values of $\lambda$ with high duty
cycles. Our model results are obviously dependent on
the value of $\lambda_c$, which changes the location of the break
in the black hole mass function, and the value of $\epsilon$, which
affects the normalization
(see SWM and Section~\ref{sec|ActiveFractions} below).

\begin{figure*}
    \includegraphics[width=16truecm]{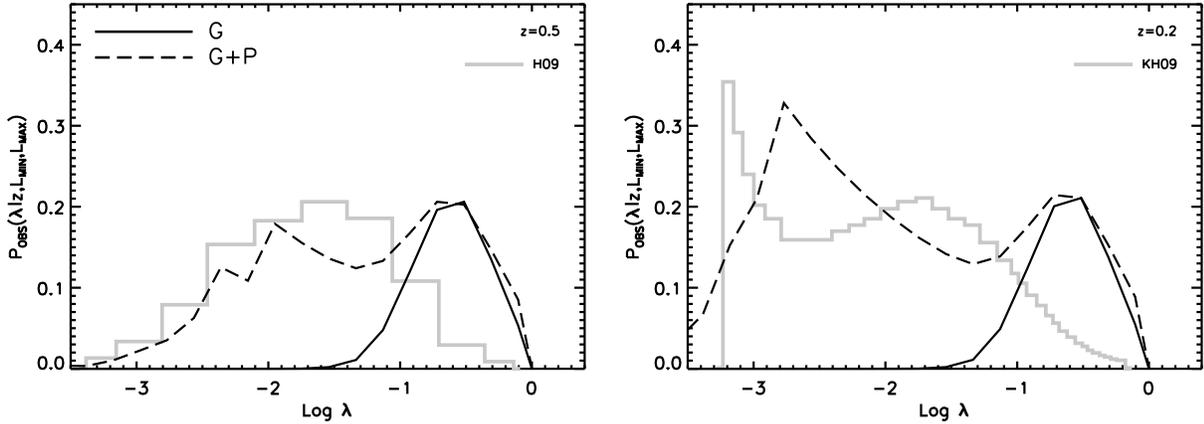}
    \caption{Eddington ratio \PL\ distributions predicted by the
    G (\emph{solid} lines) and G+P (\emph{long-dashed} lines)
    models at lower redshifts. \emph{Left panel}: Predictions
    compared to the H09 data (\emph{grey histogram})
    for black holes in the range $10^7<\mbhe/\msune<10^{10}$
    with $\log L > 43.5$ at $z=0.5$.
   \emph{Right panel}: Predictions compared to the
   KH09 data (\emph{grey histogram})
   for black holes in the range $10^7<\mbhe/\msune<10^8$ with
   $\log L> 42.5$ at $z=0.2$.}
    \label{fig|CompareKauff1}
\end{figure*}

\subsection{Comparison with measured \PL\
distributions}\label{subsec|CompareWithObservedPL}

Drawing on virial mass estimators grounded in reverberation mapping studies
(e.g., Wandel, Peterson \& Malkan 1999; Vestergaard 2002; McLure \&
Jarvis 2004)
and the availability of large samples of quasar spectra
from wide field surveys, several groups have inferred the distribution
of Eddington ratios in different ranges of redshift and luminosity.
Kollmeier et al. (2006; hereafter K06) measured \PL\ from the
AGN and Galaxy Evolution Survey (AGES; Kochanek et al. 2011), finding that
luminous AGNs at $0.5<z<3.5$ have a quite narrow range of Eddington
ratios, with a peak at $\lambda \sim 0.25$ and a dispersion
(including observational errors) of $\sim 0.3$ dex.
Netzer et al. (2007), analyzing quasar samples
from the SDSS,
found a similar result, with a slightly larger dispersion, from a sample
centred at $z\sim 2.5$. Netzer \& Trakhtenbrot (2007, see also
Vestergaard 2004; McLure \& Dunlop 2004), analyzing nearly ten
thousand SDSS quasars in the redshift range $0.3\lesssim z
\lesssim 0.75$, confirmed the log-normal shape of the Eddington ratio
distribution and also found evidence for a
significant decrease with time in the characteristic $\lambda_c$
defining the peak of the distribution.
K06 also found a factor $\sim 3$ decrease in
the peak value of $\lambda$ between high and low redshifts for
the more massive black holes in their sample.
Many other studies of the Eddington ratio distributions of quasars
and active galaxies have found
evidence for time evolution of $\lambda_c$,
and in some cases for mass-dependence
(e.g., McLure \& Dunlop 2004; Vestergaard 2004; Heckman et al. 2005;
Ballo et al. 2007; Babi\'{c} et al. 2007; Bundy et al. 2008;
Rovilos \& Georgantopoulos 2007; Cao \& Li 2008;
Fine et al. 2008; Shen et al. 2008; Greene et al. 2009; Hickox et al. 2009;
Kauffmann \& Heckman 2009; Kelly et al. 2010; Shankar et al. 2010e;
Steinhardt \& Elvis 2010a,b; Trakhtenbrot et al. 2011; Willott et al. 2010a,b; Aird et al. 2011; Shen \& Kelly 2011).
However, sample selection, observational noise, and intrinsic
scatter in black hole mass estimators can all have strong effects
on the apparent distribution of Eddington ratios
(e.g., Lamastra et al. 2006; Lauer et al. 2007; Marconi et al. 2008;
Shen et al. 2008; Netzer 2009; Shen \& Kelly 2010; Rafiee
\& Hall 2011).
At low redshifts, where samples can probe to much lower AGN luminosities,
it is clear that the distribution of Eddington ratios is much broader
than the roughly log-normal distribution found for optically luminous
quasars at high redshift.

As a set of representative examples of the literature,
we consider the Kollmeier et al. (2006; K06 hereafter) results for broad-line
optically luminous quasars at $z > 0.5$, the Hickox et al. (2009; H09
hereafter) results
at $z\approx 0.5$ for a sample selected by X-ray, infrared,
and radio emission, and Kauffmann \& Heckman's (2009; KH09 hereafter)
analysis
of the local AGN population in the SDSS, with an effective
redshift $z \approx 0.2$.

To make a close comparison of our models to
observed samples of AGNs,
we compute Eddington ratio distributions conditioned
to active AGNs with luminosities between $L_{\rm min}$ and $L_{\rm max}$, $P_{\rm obs}$, as
\begin{eqnarray}
P_{\rm obs}(\lambda|z,L_{\rm min},L_{\rm max})=\\ \nonumber
A\sum_k \sum_j
P(\lambda=L_k/(M_{\rm BH,j}l)|z,M_{\rm BH,j}) N_{\rm act}(M_{\rm BH,j},z)~,
    \label{eq|ComputePLzM}
\end{eqnarray}
where \Nact\ is defined in Eq.~\ref{eq|NactDef}.
The sums in the above equation are extended to all bins of
black hole mass, and to all bins of luminosity
in the assumed observed range, $L_{\rm min}<L_k<L_{\rm max}$.
Since we are interested here in the {\it shapes} of the predicted Eddington distributions,
the constants $A$ are chosen to renormalize the
\Pobs\ to match the peaks of the observed distributions.

Fig.~\ref{fig|CompareKollmeier1} compares
the Eddington ratio distributions predicted by the G
and G+P models to the K06 data,
for black holes in the range $10^7<\mbhe/\msune<10^{10}$
in four different bins of redshift and luminosity.
The predictions have been computed at the redshifts
of $z=1$ and $z=2$ for comparison to the $z<1.2$ and $z>1.2$ subsamples,
respectively. The predictions do not
strongly depend on these choices, since
\PL\ has no explicit redshift dependence.
Both models predict, essentially by construction, a roughly
log-normal \PL\ with a typical $\lambda_c$ that agrees with the
K06 measurements.  Most importantly, the
power-law tail of the G+P model does {\it not} lead to
substantial disagreement
because the fraction of quasars of a fixed luminosity with low values of $\lambda$ is much less than the fraction of black holes of a fixed mass with the same low values of $\lambda$.
In fact, the $\lambda$ distribution for all quasars with any
luminosity (cyan dashed curve
in the lower right panel) has the same shape as our assumed intrinsic distribution in Figure 1a.
The most significant disagreement with the data is the
high characteristic $\lambda$ for $45.5 < \log L < 46$ at $z<1.2$
(lower left panel), a consequence of keeping the model $\lambda_c$
fixed instead of allowing redshift evolution.

Fig.~\ref{fig|CompareKauff1} compares the predicted and observed
$\lambda$ distributions for
the lower redshift samples of H09 at
$z\approx 0.5$ and KH09 at $z\approx 0.2$.
To roughly take into account the effective limits in black hole mass
and luminosity of the observational samples,
for the former we consider all black holes with $\mbhe > 10^7\, \msune$
shining at $L > 10^{43}\ergse$, while in the latter case we
adopt the restricted range $10^7 < \mbhe/\msune < 10^8$
shining above $L=10^{42.5}\, \ergse$.
The match to the observations is poor in both cases.
The distributions predicted by the G model are much narrower than
the observed distributions, and they peak at a value
$\lambda_c \approx 0.25$ that is roughly ten times
higher than the high-$\lambda$ peaks of the observed
distributions.
The power-law tail in the G+P model is in rough agreement
with the observed distributions at low $\lambda$,
but the peak of the Gaussian component remains discrepant.
Fitting these observations, even approximately, requires evolution
of $\lambda_c$, decreasing towards low redshift.
We return to this point in Section~\ref{subsec|EddingtonRatioDistributions}
below, after first finding independent evidence for evolution
in $\lambda_c$ from observed active galaxy fractions.

\begin{figure*}
    \includegraphics[width=17truecm]{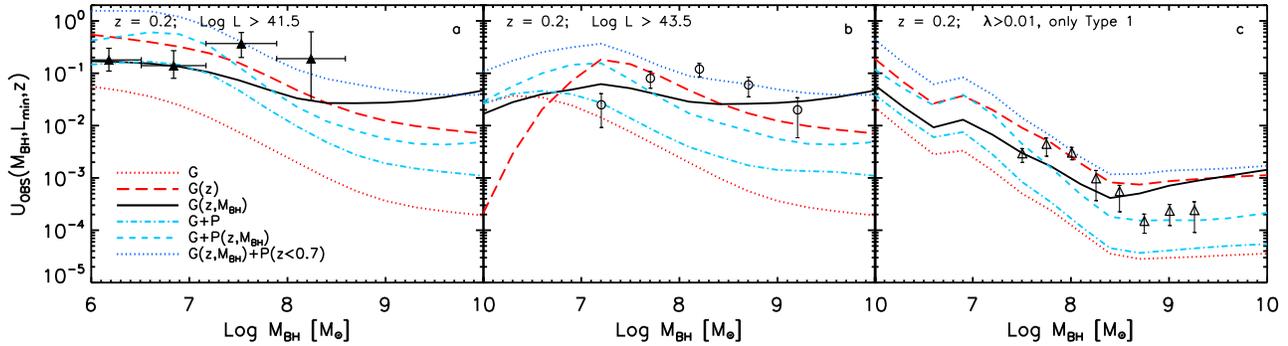}
    \caption{
    Comparison of predicted and observationally estimated duty
    cycles at low redshift.
    \emph{Left panel}: Data points from Goulding et al.\ (2010)
    compared to model predictions at $z=0.2$ as labelled,
    with a luminosity threshold $\log L >41.5$.
    \emph{Middle panel}: Data points from Kauffmann et al. (2003)
    compared to model predictions, with a luminosity threshold
    $\log L > 43.5$.
    \emph{Right panel}: Data points from Schulze \& Wisotzki
    (2009) compared to model predictions, with an {\it Eddington ratio} (not luminosity) threshold $\lambda>0.01$
    and a correction factor for the Type 1 Broad Line AGNs taken from Greene \& Ho (2009).  The three
    lowest mass points include the Schulze \& Wisotzki (2010)
    incompleteness corrections, which are negligible at
    higher mass.
    }
    \label{fig|DutyCyclez0}
\end{figure*}

\section{Active Galaxy Fractions: redshift and mass-dependence of
\PL}\label{sec|ActiveFractions}

Duty cycles are a basic prediction of continuity equation
models as discussed in Section~\ref{sec|formalism}.
If we assume that most or all massive galaxies contain black holes,
as supported by observations of the local universe (see, e.g., Ferrarese \& Ford 2005, and references therein), then
active galaxy fractions can be used as an observable proxy for
black hole duty cycles.
However, active galaxy fractions depend on the
luminosity or Eddington ratio threshold of each specific
observational sample:
lower thresholds obviously increase the
active fraction. Therefore, in order to properly compare model predictions
to available data sets, we
define the fraction of black holes shining above a luminosity
threshold $L_{\rm min}$, or observed duty cycle $U_{\rm obs}$, as
\begin{eqnarray}
U_{\rm obs}(\mbhe,L_{\rm min},z)=\int_{\log \lambda_{\rm min}}^{\infty}
P(\lambda|\mbhe,z) U(\mbhe,z)d \log \lambda\, , \\
\lambda_{\rm min} = L_{\rm min}/(l \mbhe)\, .  \nonumber
    \label{eq|DutyCycleAboveL}
\end{eqnarray}

In this Section and Section~\ref{sec|OtherConstraints} we will
introduce versions of our reference models that incorporate redshift
and mass dependence of $P(\lambda)$, and we will consider a variety
of ``test'' models that illustrate specific points.  The key features
of these models are summarized in Table~\ref{table|models}, and we
will summarize the qualitative successes and failures of the
reference models in Table~\ref{table|models2} below (Section~\ref{sec|Discussion}).
These model changes generally have little impact on the overall mass
accretion histories, so we will not repeat the analysis shown previously
in Figures~\ref{fig|BHMFandDuty} and~\ref{fig|RhoBHz} but instead focus
on predictions where the new models differ significantly from those
described in Section~\ref{sec|broadPL}.  Despite systematic uncertainties
in the observational constraints, even qualitative trends with
redshift and black hole mass are enough to provide strong model tests.

\subsection{Observational Estimates}\label{subsec|ObservationalEstimates}

We first summarize the observational estimates of AGN fractions
that we adopt for our model comparisons, with further details of
some of these estimates given in
Appendix~\ref{Appendix|AGNdutyCycles}.
For high luminosity, broad-line (Type 1) AGN in the local Universe,
Schulze \& Wisotzki (2010) have estimated the mass function of
active black holes at $z< 0.3$ by applying linewidth mass estimators
to quasars identified in the Hamburg/ESO objective prism survey.
They divide their active black hole density by the total black
hole space density in the Marconi et al.\ (2004) mass function to
derive an active fraction.  Schulze \& Wisotzki impose a threshold
of $\lambda \geq 0.01$ in their estimated Eddington ratio to define
active systems.  Host galaxy contamination causes incompleteness
in their AGN catalog below $\mbhe \approx 10^{7.5}M_\odot$
(for $\lambda \geq 0.01$).  In the right panel of
Fig.~\ref{fig|DutyCyclez0} we plot the points from their
Fig.~12, correcting their three lowest $\mbhe$ points for
incompleteness as suggested by the authors.
According to their simulations, higher $\mbhe$ points should be unaffected by incompleteness,
and we have omitted points below $\mbhe = 10^7 M_\odot$ where
incompleteness corrections become large.
The active black hole fractions estimated by Schulze \& Wisotzki (2010)
are low, only $\sim 2 \times 10^{-4}$ at $\mbhe = 10^9 M_\odot$,
rising to $\sim 10^{-3}$ at $\mbhe = 10^8 M_\odot$.

\begin{figure*}
    \includegraphics[width=17truecm]{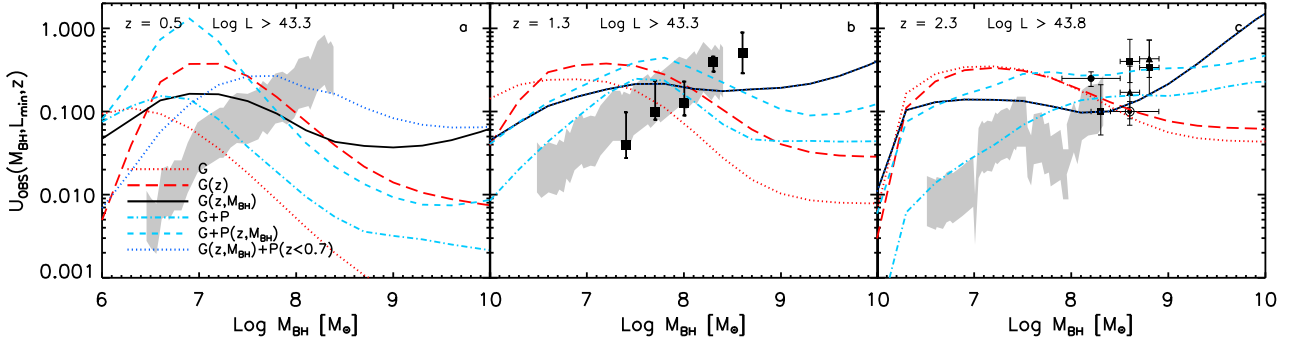}
    \caption{Predicted and observationally estimated duty cycles
    at redshifts $z=0.5$, $z=1.3$, and $z=2.3$.
    In all panels, \emph{grey bands} show the estimates of
    Xue et al. (2010).
    Points in the \emph{middle} panel, points are from Bundy et al.\ (2008).
    Points in the \emph{right} panel come from
    Erb et al. (2006; \emph{filled small squares}), Kriek
    et al. (2007; \emph{filled triangles}), Alexander et al. (2005; 2008;
    \emph{filled circle}), and Caputi et al. (2006; \emph{semi-filled circle}).
    }
    \label{fig|DutyCycleHighz}
\end{figure*}

Kauffmann et al.\ (2003) estimate much higher active fractions
for narrow line (Type 2) AGN identified by applying emission
line diagnostics to SDSS spectra in the SDSS main galaxy redshift
survey.  Since Kauffmann et al.\ (2003) begin with a galaxy catalog,
quasars that are bright enough to appear as point sources (rather than
extended sources) in SDSS imaging are excluded.
They estimate their completeness limit at $10^7\, L_\odot$ in
[OIII] luminosity, which we convert to an approximate
$L_{\rm min} = 10^{43.5}\ergse$ by adopting an extinction-corrected
bolometric correction of $\sim 800$ (both Kauffmann et al.\ 2003 and KH09
adopt extinction-corrected luminosities).
Points in the middle panel of Fig.~\ref{fig|DutyCyclez0}
are taken from Fig.~5 of Kauffmann et al.\ (2003),
where we have simply converted stellar mass to black hole
mass by multiplying by $1.6\times 10^{-3}$ (Magorrian et al.\ 1998).
This crude conversion should be reasonably accurate above
$M_* \sim 10^{11.5} M_\odot$, but it probably overestimates
the black hole mass for lower mass galaxies that are no longer
bulge-dominated.  The Kauffmann et al. (2003) active fractions
rise from $\sim 2\%$ at $\mbhe = 10^9 M_\odot$ to
$\sim 10\%$ at $\mbhe = 10^8 M_\odot$, then decline towards
lower $\mbhe$.  The $10^{43.5}\ergse$ luminosity threshold
corresponds to $\lambda = 0.01$ at $\mbhe \approx 10^{7.5} M_\odot$,
so at higher masses the Kauffmann et al.\ (2003) estimates correspond
to a lower $\lambda$-threshold than Schulze \& Wisotzki's.

The $1.5$ to 2 order-of-magnitude gap between the Kauffmann et al.\ (2003)
and Schulze \& Wisotzki (2009) active fractions
can be partly explained by the different abundances of
broad line and narrow line AGN at low redshift. The abundance ratio of
the two types of AGN has been addressed by
Greene \& Ho (2009; a correction of Greene \& Ho 2007), who
analyze all SDSS spectroscopic objects (targeted as galaxies
or quasars) in Data Release 4 (Adelman-McCarthy et al.\ 2006)
to define a sample of 8,400 broad-line AGN at $z \la 0.3$.
Their estimated active fractions (in their Fig.~11) are
close to those of Schulze \& Wisotzki. They find a gap
of 0.9-1.3 dex between their active black hole mass function for
broad-line AGN and the corresponding result (Heckman et al. 2004)
for SDSS narrow-line AGN (see Greene \& Ho 2009, Figure 10),
which implies a correction of $\sim 10-20$ for the ratio of
Type I to Type II AGN at low redshift.
The remaining factor of $5-10$ may be accounted for by
the effects of scatter in bolometric corrections,
different luminosity thresholds, and systematic offsets in the
black hole mass function estimates.
Note that scatter in $L_{\rm bol}/L_{\rm [OIII]}$ could
make the effective luminosity threshold of the Kauffmann et al. (2003)
data lower than we have assumed, in particular if the luminosity function
is affected by a steeply rising low-$\lambda$ tail in $P(\lambda)$.

For studies that probe to lower AGN luminosities, active galaxy
fractions are even higher than those of Kauffmann et al. (2003).
For example, Ho (2004) argues that at least 40\% of local
bulge-dominated galaxies host a low luminosity AGN and/or a LINER,
and Grier et al. (2011) find nuclear X-ray sources in $\sim 60\%$
of galaxies in the SINGS survey (Kennicutt et al. 2003).
As a representative example of low luminosity statistical studies,
we show in the left panel of Fig.~\ref{fig|DutyCyclez0} the data
of Goulding et al. (2010, their Fig.~5), based on an X-ray and IR
census of a volume-limited sample of galaxies with $D < 15$ Mpc.
Their active fractions range from 14\% to 37\% over the black
hole mass range $10^6 - 10^{8.5}\, \msune$, albeit with large
statistical errors reflecting the small sample size.
The luminosity threshold of this sample is very low,
about $L_{\rm min} = 10^{41.5}\, \ergse$ in bolometric units (their
Fig.~3).

At higher redshifts we take as our primary data set the measurements
reported by Xue et al.\ (2010) based on moderate luminosity X-ray
AGN in the Chandra Deep Fields.  We reproduce results from their
Fig.~14 (the upper, orange bands) as grey bands in the three
panels of Fig.~\ref{fig|DutyCycleHighz}, converting their
X-ray luminosity thresholds to bolometric luminosity thresholds
with our luminosity-dependent bolometric correction (Section~\ref{subsec|AGNLF}).
We again convert stellar
masses to black hole masses by simply multiplying by the
Magorrian et al. (1998) factor of $1.6\times 10^{-3}$.  This scaling should
be taken with a grain of salt, as we are using total stellar masses
rather than bulge masses and ignoring possible redshift evolution
of the scaling factor.  However, these uncertainties should not
affect the key lessons that we take from Xue et al.\ (2010): at all
three redshifts, the duty cycle for BHs shining above $\log L>43.3$
increases with increasing $\mbhe$
(indicated in the data by increasing $M_*$), and for the most massive
galaxies active fractions are $10-50\%$ (grey bands marking the 1-$\sigma$ uncertainty).

Other points in Fig.~\ref{fig|DutyCycleHighz} come from other
studies with similar redshift ranges and luminosity thresholds.
At intermediate redshifts, the data of Bundy et al.\ (2008)
from the DEEP2 and AEGIS surveys show nearly the same
mass trend and normalization as Xue et al. (2010). More recently,
Mainieri et al. (2011) also found evidence for a steep
increase in AGN activity with stellar mass at $\langle z \rangle \approx 1.1$,
though with a luminosity threshold about two orders
of magnitude higher than the one by Xue et al. (2010).
At $z>2$ we add several points for high mass black holes
based on the studies of Caputi et al. (2006), Erb et al. (2006),
Kriek et al. (2007), and Alexander et al. (2005, 2008).
These data have been collected at bolometric
luminosities approximately $\log L > 43.8$, comparable to those of Xue et al. (2010).
Further details about these observations are discussed in
Appendix~\ref{Appendix|AGNdutyCycles}.
They extend the mass trend found by Xue et al. (2010), with
active fractions ranging from $\sim 10\%$ to $\sim 50\%$.

\subsection{Low-redshift AGN fractions: A redshift-dependent \PL?}\label{subsec|PLzdependence}

Red dotted and cyan dot-dashed curves in Fig.~\ref{fig|DutyCyclez0} show
the duty cycles \UM\ predicted by the G and G+P models at $z=0.2$,
where we have imposed the bolometric luminosity thresholds
$\log L > 41.5$ (left panel), $\log L > 43.5$ (middle), and
$\lambda > 0.01$ (right).  For the right panel, we have additionally
multiplied model predictions by a mass-dependent factor
(ranging from $\sim$ 15\% at $\mbhe = 10^7 \msune$ to $\sim 3\%$ at $\mbhe = 3 \times 10^8 \msune$)
to account for Greene \& Ho's (2009) estimate of broad line AGN fractions.
(Specifically, the correction is obtained from the ratio of Broad Line to
Type II number densities
in their Figure 10, with an extrapolation of their results
above $\mbhe \sim 5 \times 10^8 \msune$.)
Both models drastically underpredict all measured active fractions for $\mbhe
> 10^{7.5} M_\odot$.
Predicted duty cycles are higher for the G+P model, and the gap
between G+P and G is larger for lower luminosity thresholds that allow
a larger contribution of sub-Eddington black holes, but the G+P model
still falls well short of the observed active fractions.

As introduced by Shankar et al. (2004) and discussed more
extensively by SWM (see their Figure 11), one can increase
predicted duty cycles at low redshift by adopting a redshift-dependent
Eddington ratio distribution, with characteristic $\lambda_c$
that drops towards low redshifts.  Decreasing $\lambda_c \propto L/\mbhe$
maps a given luminosity to more massive, and hence rarer, black holes,
thus requiring a higher duty cycle to reproduce the observed AGN
space density.  Furthermore, the results of H09,
KH09, and other studies directly suggest a
lower $\lambda_c$ at low redshifts, as already discussed in
Section~\ref{subsec|CompareWithObservedPL} (see Fig.~\ref{fig|CompareKauff1}).
Motivated by these results, we introduce models in which the location
of the peak in the G and G+P \PL\ distributions decreases
from $\lambda_c=1.0$ at $z=6$ to $\lambda_c \sim 0.02$ at $z\sim 0.1$,
following
\begin{equation}
\lambda_c(z)=\lambda(z=6)\left[\frac{1+z}{7} \right]^{\alpha}\, ,
\label{eq|PLz}
\end{equation}
with $\alpha=2.2$, although we note that even lower values of $\alpha$
yield similar results.
Eq.~\ref{eq|PLz} implies $\lambda_c$ = 0.71, 0.48, 0.29, 0.16, 0.06,
0.03, 0.02 at $z=5$, 4, 3, 2, 1, 0.5, and 0.1, respectively.

In agreement with SWM,
who considered $\delta$-function \PL, we find that this
evolution has a minor impact on the evolved black hole mass function (at
least at high masses) or on
the global accretion histories
shown in Fig.~\ref{fig|RhoBHz}. However,
predicted duty cycles at $z\sim 0.2$ for the \Gz\ model ---
a Gaussian \PL\ with the $\lambda_c(z)$ given by
Eq.~\ref{eq|PLz} --- are much higher than those of the
G model, as shown by comparing
the dotted and long-dashed curves in
Fig.~\ref{fig|DutyCyclez0}.  The model now approximately agrees with the
Goulding et al. (2010), Kauffmann et al. (2003), and Schulze \& Wisotzki
(2010) active fractions,
though at high masses it is above the latter data and below
the former two.
Solid and short-dashed curves in Fig.~\ref{fig|DutyCyclez0} show
models with Gaussian and G+P $\lambda$-distributions
that incorporate both redshift and mass dependence of $\lambda_c$,
as discussed in the next section.

The redshift dependence expressed in Eq.~\ref{eq|PLz} is by
no means unique.
In particular, we have checked that an
equally good match to the local data can be obtained by allowing
the characteristic \lamc\
to decrease only below redshift $z\sim 1$, while remaining close to unity
at higher redshifts.
This abrupt change in \lamc\ at late times\footnote{We note that Shankar, Bernardi \& Haiman (2009) found that a constant
Eddington ratio as a function of redshift was consistent with the
statistics of local, early-type galaxies coupled with a mild evolution in the black hole-velocity dispersion scaling relation and the integrated AGN energy density.
However, their conclusions
were driven by assuming a non-evolving structural evolution of their hosts.
Allowing for some evolution in velocity dispersion (as suggested by several observations
and models, as in, e.g., Shankar et al. 2011, and references therein)
would be consistent with a decrease in time of the characteristic Eddington ratio
while preserving the mild evolution in the black hole-velocity dispersion relation.} leads to duty cycles that increase
at fixed black hole
mass from $z\sim 1$ to $z\sim 0$, instead
of decreasing as predicted by continuous redshift evolution.
This increasing behaviour may be in conflict with observations;
some direct X-ray
and optical analyses show that the fraction
of active galaxies above a fixed stellar mass decreases with decreasing
redshift in the range $0<z<1$ (e.g., Shi et al. 2008).

\subsection{AGN fractions vs. mass: A mass-dependent \PL?}
\label{subsec|PLmassdependence}

Fig.~\ref{fig|DutyCycleHighz} compares the model predictions to
the higher redshift active fraction data discussed in
Section~\ref{subsec|ObservationalEstimates}.  The G and \Gz\
models both predict a duty cycle that declines with increasing $\mbhe$,
opposite to the trend found by
Xue et al.\ (2010) and Bundy et al. (2008).
At first glance, the G+P model appears to fare
better, at least in the $z=1.3$ and $z=2.3$ panels.
However, we view this better agreement as artificial ---
it is a consequence of the high initial black hole space
density required to keep high-$z$ duty cycles below
unity in this model (see Fig.~\ref{fig|BHMFandDuty}).
The high black hole space density inherited from these initial
conditions leads to a lower duty cycle for low mass black holes.
We regard these high initial space densities as observationally
untenable for the reasons discussed in Section~\ref{subsec|AccretionHistories},
so we discount this apparent success of the G+P model,
which vanishes by $z=0.5$ in any case.

\begin{figure*}
    \includegraphics[width=15truecm]{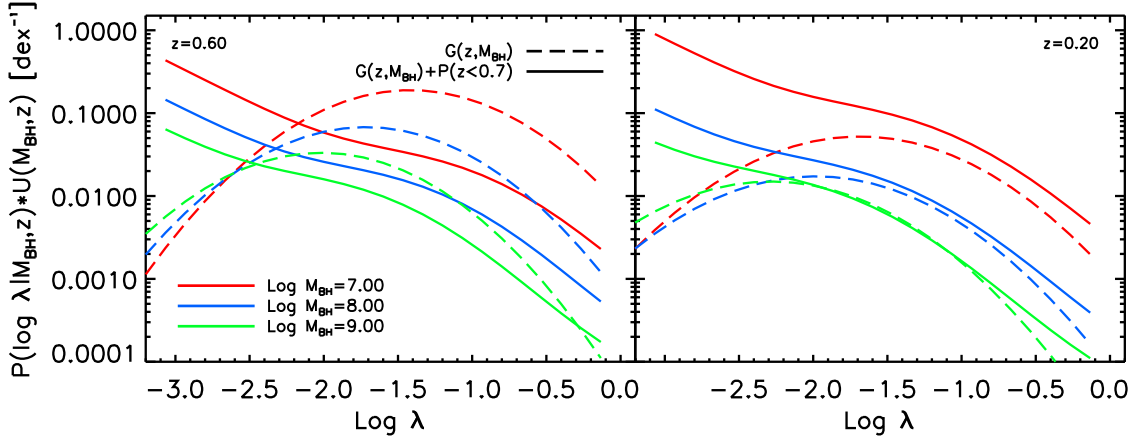}
    \caption{\PL\ distributions for our \GzM\ model ({\it dashed} curves)
    and the \GzM\ model with a steep power-law component added
    at low redshift ({\it solid} curves), shown at $z=0.6$ ({\it left})
    and $z=0.2$ ({\it right}).  In each panel, the three curves
    show \PL\ for black hole masses of $10^7 M_\odot$,
    $10^8 M_\odot$, and $10^9 M_\odot$ from top to bottom, as labelled.}
\label{fig|SteepPowerLaw}
\end{figure*}

The only other way we have found to make the predicted duty cycle
increase with black hole mass at these redshifts is to introduce
an explicit mass dependence of \PL, with $\lambda_c$ declining
with increasing mass.  In this case massive black holes
are matched to more common, lower luminosity AGN, implying
a higher duty cycle.  There is some direct empirical support for
such a trend (e.g.  Heckman et al. 2004; Ballo et al. 2007;
Babi\'{c} et al. 2007; Netzer \& Trakhtenbrot 2007;
Rovilos \& Georgantopoulos 2007; Fine et al. 2008; H09),
and it could arise theoretically in models that envisage
a faster shut-off of activity in more massive systems due to AGN feedback
(e.g., Matteucci 1994; Granato et al. 2006; Lapi et al. 2006).
Motivated roughly by these empirical studies, we have adopted
a model with the same redshift dependence as before and a
mass dependence
\begin{equation}
\log \lambda_c(\mbhe,z)=\log \lambda_c (z) +\left [\beta_M (\log \mbhe-C)
\right] \, ,
\label{eq|PLmassDependent}
\end{equation}
with $\beta_M=-0.3$ and $C=7$.
Eq.~\ref{eq|PLmassDependent} implies a $\lambda_c$ that drops by
a factor of two between $10^7 M_\odot$ and $10^8 M_\odot$ and
by a further factor of two between $10^8 M_\odot$ and $10^9 M_\odot$.

Solid lines in Figures~\ref{fig|DutyCyclez0} and~\ref{fig|DutyCycleHighz}
show predictions of the \GzM\ model, which incorporates mass-dependence
of the radiative efficiency
$\epsilon$ as well as of $\lambda_c$, for reasons that we will
discuss shortly. This model also incorporates a broadening of the
Gaussian in the \GzM\ at lower redshifts following the trend
\begin{equation}
\log \Sigma_\lambda(z)=\log \Sigma_\lambda(z=6)-0.4\log
\left(\frac{1+z}{7}\right)  \, ,
\label{eq|Sigma}
\end{equation}
with $\log \Sigma_\lambda(z=6)=0.3$, yielding $\Sigma_\lambda(z) \sim 0.4,
0.5, 0.6$ at $z=3$, 2, 0.2.
As discussed below, we include this additional modification to the model to
provide a better match
to the local Eddington ratio distributions.
At all redshifts, this model predicts a duty
cycle that is roughly flat or rising
with black hole mass, though the rising trend is always
weaker than that found by Xue et al.\ (2010) and Bundy et al.\ (2008).
We could adopt a still stronger mass dependence in this model,
which would improve agreement with the observed trends,
but this would no longer be supported by empirical estimates
(e.g., Hickox et al.\ 2009), and it would exacerbate problems in
explaining the Eddington ratio distributions and local black hole mass function
(see below).
The \GzM\ model is also in rough agreement with all
the local data sets on active fractions (solid lines in
Fig.~\ref{fig|DutyCyclez0}), though it overpredicts the Schulze \&
Wisotzki (2010) active fractions for $\mbhe \sim 10^9\, \msune$
by a factor of several.

Short-dashed lines in
Fig.~\ref{fig|DutyCycleHighz} show the predictions of
a model labeled \GPzM\ that has a G+P form with redshift-
and mass-dependent $\lambda_c$ and mass-dependent $\epsilon$
like that of the \GzM\ model.  Instead of allowing the Gaussian
to broaden at low redshift via Eq.~\ref{eq|Sigma}, we keep its
width fixed at $\Sigma_\lambda = 0.3$ but allow the power-law
tail to grow over time.
Specifically, the $P(\lambda)$ distribution is cut off at
\begin{equation}
\label{eq|LambdaMin}
\lambda_{\rm min}(z)= 0.1\left({1+z \over 7}\right)^3~,
\end{equation}
implying $\lambda_{\rm min} = $ 0.1, 0.063, 0.036, 0.019,
0.0079, 0.0023, 0.0098, 0.00039 at
$z = 6$, 5, 4, 3, 2, 1, 0.5, 0.1.
This simple modification (similar to that adopted by Cao 2010)
avoids the need for a large initial black hole seed
population, thus removing the artificial aspect of
our redshift-independent G+P model, but it produces
a full G+P distribution at intermediate and low redshifts.
The predictions of this model are similar to those of \GzM,
with a somewhat worse match to the $z=1.3$ data in
Fig.~\ref{fig|DutyCycleHighz}b and a somewhat better
match to the Schulze \& Wisotzki data in
Fig.~\ref{fig|DutyCyclez0}.  Both models fail to
reproduce the rising $U(M_{\rm BH})$ found at
$z=0.5$ by Xue et al. (2010).

\subsection{AGN fractions vs. mass: a steep power-law component at low $z$?}
\label{subsec|PLmassdependence2}

At early stages of our investigation, we anticipated that the observed
trends of rising active fraction with rising galaxy mass might be
explained mainly by the combination of a broad $P(\lambda)$ with
sample luminosity thresholds: high mass black holes remain observable
at lower $\lambda$ values, so they could have higher duty cycles
above fixed luminosity even if the trend of total duty cycle
(over all $\lambda$ values) was flat or decreasing.
This is essentially the explanation advanced by
Aird et al.\ (2011), who fit AGN data from the
PRIMUS survey with a model in which all black holes have
an Eddington ratio distribution $P(\lambda) = P_0(z)\times (\lambda/\lambda_0)^{-0.7}$
independent of mass.

As already discussed in the context of our
G+P model, we find that a long power-law tail is unrealistic
at high redshift because it implies duty cycles higher than unity
unless the population of seed black holes is implausibly large.
However, a power-law component that kicks in at lower redshifts
is still feasible. The blue dotted lines in
Figures~\ref{fig|DutyCyclez0} and \ref{fig|DutyCycleHighz}
show the predicted duty cycles for a model equivalent to
\GzM\ with the addition
of a power-law component at $z < 0.7$ with slope $\alpha=-0.9$
and normalized to have the same value as the Gaussian at $\log \lambda=\log \lambda_c-0.2\log \Sigma_{\lambda}$.
Fig.~\ref{fig|SteepPowerLaw} plots \PL\ for this model at
three different black hole masses and two redshifts, in comparison
to those of the \GzM\ model.  For this plot we have multiplied \PL\
by \UM\ so that the curves are normalized to the duty cycle rather
than integrating to unit probability.  The Gaussian and power-law
components join to produce a \PL\ that is close to a single
power law from $\lambda = 10^{-3}$ to $\lambda=1$.

This model still exhibits ``downsizing'' in the sense that higher
mass black holes have lower duty cycle at any given $\lambda$.
However, the duty cycle for AGN active above a luminosity
threshold $\log L = 43$ {\it increases} with black hole mass,
as shown in the left panel of Fig.~\ref{fig|DutyCycleHighz},
because massive black holes can shine above the threshold
at low $\lambda$.  This is the only model we have constructed
that reproduces the Xue et al.\ (2010) trend at $z=0.5$.
The Aird et al.\ (2011) prescription, with the \PL\ normalization
independent of mass, would produce a still stronger rising trend,
but we find that such a model cannot simultaneously match our
evolved black hole mass function and our input luminosity function.

\section{Other observational constraints}
\label{sec|OtherConstraints}

\subsection{The local mass function: A mass-dependent radiative efficiency?}
\label{subsec|epsilonmass}

\begin{figure*}
    \includegraphics[width=15truecm]{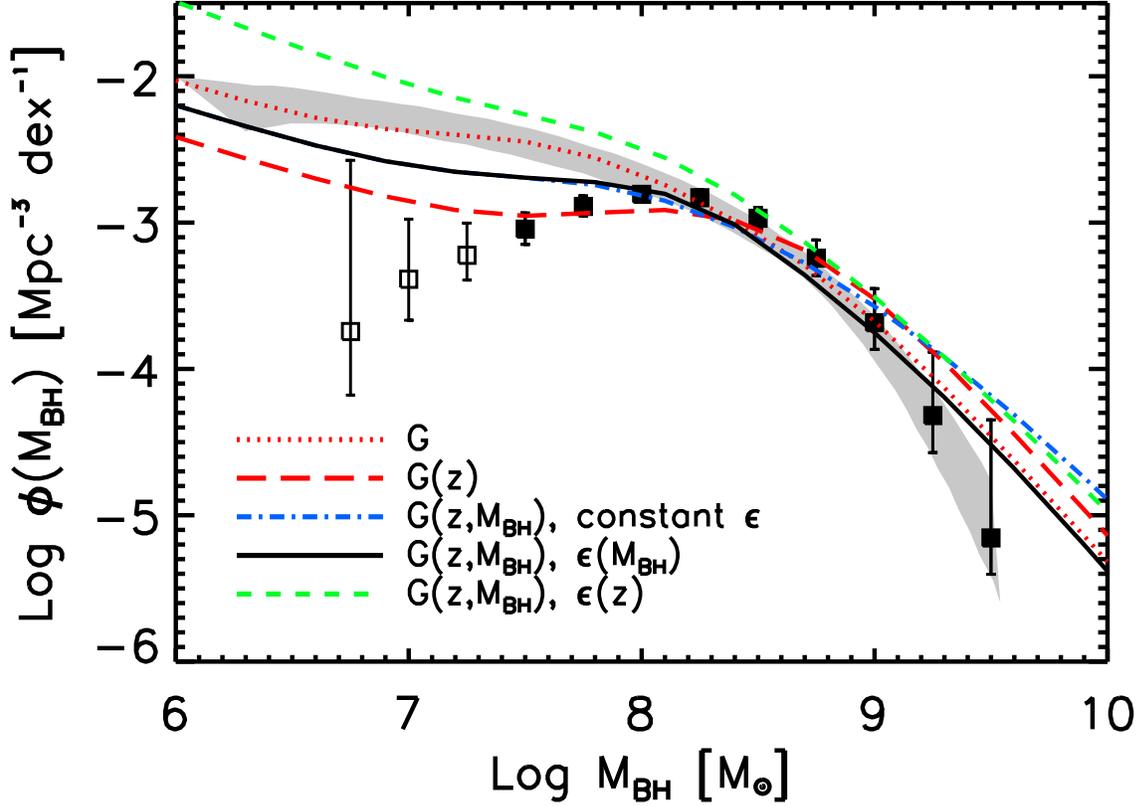}
    \caption{Local black hole mass function
    predicted by different models, as labelled.
    Data are as in Fig.~\ref{fig|BHMFandDuty}.
    Of the three \GzM\ models, with mass-dependent $\lambda_c$,
    only the one with mass-dependent $\epsilon$ (solid line) matches
    the high end of the mass function.  Elsewhere in the paper
    this preferred model is simply labeled \GzM.}
    \label{fig|BHMFcomparison}
\end{figure*}

As shown previously in Fig.~\ref{fig|BHMFandDuty}, our simple, non-evolving
$\delta$, G, and G+P models yield reasonable agreement with
SWM's estimate of the local black hole mass function for
$\mbhe \leq 10^9 M_\odot$, but the model predictions skirt
the upper boundary of the observational estimates at higher
masses.  Fig.~\ref{fig|BHMFcomparison} plots mass functions
for the Gaussian \PL\ models with various assumptions about
redshift evolution and mass-dependence.
Redshift evolution alone (red long-dashed line) produces modest changes above
$\sim 2 \times 10^8 M_\odot$, but the decreasing $\lambda_c$ boosts growth
of high mass black holes at the expense of low mass black
holes, leading to a flattening of $\Phi(\mbhe)$ near
$10^8 M_\odot$ and a lower space density of lower mass black holes.
This change partly reproduces the turnover found by
Vika et al. (2009), though it still does not reproduce
their estimate below $\mbhe \sim 10^{7.5} M_\odot$.

Adding only the mass-dependent $\lambda_c$ of Eq.~\ref{eq|PLmassDependent}
leads to the ``\GzM, constant $\epsilon$'' curve (blue dot-dashed line)
in Fig.~\ref{fig|BHMFcomparison}.
For this model, we have lowered the radiative efficiency from
$\epsilon = 0.06$ to $\epsilon = 0.05$ to improve the match to the amplitude
of $\Phi(\mbhe)$ at $10^8-10^{8.5} M_\odot$, where it is
best measured.  However, the model then strongly overpredicts
the mass function at $\mbhe > 10^9 M_\odot$, a direct
consequence of the higher duty cycles of high mass black holes
that the model was designed to produce.  As discussed in
Section~\ref{subsec|Mergers} below, the expected impact of black
hole mergers would make this overprediction even more severe.

Volonteri, Sikora \& Lasota (2007), Cao \& Li (2008), and Fanidakis et
al. (2011), among others, have suggested that
radiative efficiency may
increase with black hole mass, perhaps mirroring a merger-induced
increase in the average spin for the more massive black
holes. Cao \& Li (2008) claimed evidence for an increasing
\epsi\ from the match between the predicted and local black
hole mass functions at the high-mass end. More recently
Davis \& Laor (2011) directly determined the radiative efficiency
from the ratio between bolometric luminosity and accretion rate, the latter
determined from thin accretion disk model fits to the optical luminosity
density.
Their analysis seems to support an increase of the radiative efficiency
with black hole mass, from 0.03 at low masses to 0.4 at high masses.
An increasing radiative efficiency at high masses
is also consistent with the notion that the bulk of luminous radio
AGNs, believed
to be rapidly spinning black holes, are indeed massive black holes (e.g.,
McLure \& Jarvis 2004; Metcalf \& Magliocchetti 2006;
Shankar et al. 2008a,b; Shankar et al. 2010e).

Inspired by these theoretical and empirical arguments,
we have constructed the \GzM\ model (see Tables~\ref{table|models} and \ref{table|models2}), which, following Cao \& Li (2008), adopts a
mass-dependent \epsi\ given by:
\begin{equation}
\epsilon=\left\{
  \begin{array}{ll}
    0.05 & \hbox{if $\mbhe < 10^8 $\msun} \\
    0.05(M_{\rm BH}/10^8\, M_{\odot})^{0.3} & \hbox{if $\mbhe \ge 10^8
    $\msun\,\, .}
  \end{array}
\right.  \label{eq|radefficiencyMass}
\end{equation}
This change reduces the implied accretion rates of massive black
holes, leading to much better agreement with the local
$\Phi(\mbhe)$ as shown by the solid curve in
Fig.~\ref{fig|BHMFcomparison}. This model also has a stronger
positive trend of duty cycle with $\mbhe$, improving agreement
with the data in Fig.~\ref{fig|DutyCycleHighz}.
The two improvements are connected: raising $\epsilon$ decreases
the space density of massive black holes, so a higher duty cycle
is required to match the space density of active black holes.
The mass function (not shown) of the model introduced in
Section~\ref{subsec|PLmassdependence2}, with a steep power-law
at $z<0.7$, is nearly identical to that of \GzM, except at
$\log M \leq 7.2$, where it is lower by $\sim 0.3$ dex.

Wang et al. (2009) have instead recently claimed empirical
evidence for an increase of \epsi\ with redshift.
Their results are based on an inversion of Soltan's (1982) argument,
expressing the radiative efficiency at any redshift
as the ratio of the accreted mass density up to that redshift
$\rho_{\rm BH}(z)$ to the
corresponding total emissivity obtained by direct
integration of the AGN luminosity function (see their Eq.~6).
The mass density $\rho_{\rm BH}(z)$ was computed by first measuring the
mass density locked up in all active
black holes at $z$ (with masses from virial relations), then correcting
by the duty
cycle extracted from the number counts of active galaxies in VIMOS-VLT
Deep Surveys.
Clustering analysis also provides hints of redshift-dependent
radiative efficiency.
Shankar et al. (2010b), adopting basic
accretion models and cumulative number matching
arguments, found that black hole accretion plus merger models consistent
with both the quasar luminosity function and the
strong observed clustering at $z \approx 4$ (Shen et al. 2007)
must be characterized by high duty
cycles and large radiative efficiencies $\epsilon \gtrsim 0.2$, if they
are accreting at a significant fraction of the Eddington limit.
However, an efficiency $\epsilon \geq 0.2$ at all redshifts
would underpredict the local black hole mass function (SWM).
In flux-limited quasar surveys, higher redshift quasars are found
above higher luminosity thresholds, so
both of these empirical arguments could potentially be answered by a
mass-dependence of the sort implied by Eq.~\ref{eq|PLmassDependent}.

We nonetheless consider a directly redshift-dependent model,
approximating these empirical findings with the relation
\begin{equation}
\epsilon(z)=0.0022\left[1+{\rm erfc} \left(-\frac{z}{1.5}\right)^6\right]\, .
\label{eq|radEfficiencyReds}
\end{equation}
Eq.~(\ref{eq|radEfficiencyReds}) implies
$\epsilon \approx 0.14$, 0.14, 0.12, 0.05, 0.02, 0.01
at $z=6$, 4, 2, 1, 0.5, 0.2, respectively.
The local black hole mass function predicted by this G$(z,\mbhe)+\epsilon(z)$
model,
shown by a green dashed curve in Fig.~\ref{fig|BHMFcomparison},
tends to overestimate the observed mass function at all scales.
The match could be improved
by increasing the overall normalization of $\epsilon(z)$, still
allowed (and actually preferred) by the measurements of Wang et al. (2009;
see their Figure 2\emph{a}), but such an increase
produces unphysical models with duty cycles significantly
higher than unity, and it still leaves too many high mass black holes.
Moreover, the cosmological accretion rate predicted by the
G$(z,\mbhe)+\epsilon(z)$ model is morphologically different from the
cosmological star formation rate of galaxies, at variance
with the agreement shown in Fig.~\ref{fig|RhoBHz}.
In itself this is not a fatal objection, but it then requires
non-trivial fine tuning to reproduce a
tight local relation between black hole mass and stellar mass.

\subsection{Eddington Ratio Distributions, Revisited}
\label{subsec|EddingtonRatioDistributions}

\begin{figure*}
    \includegraphics[width=15truecm]{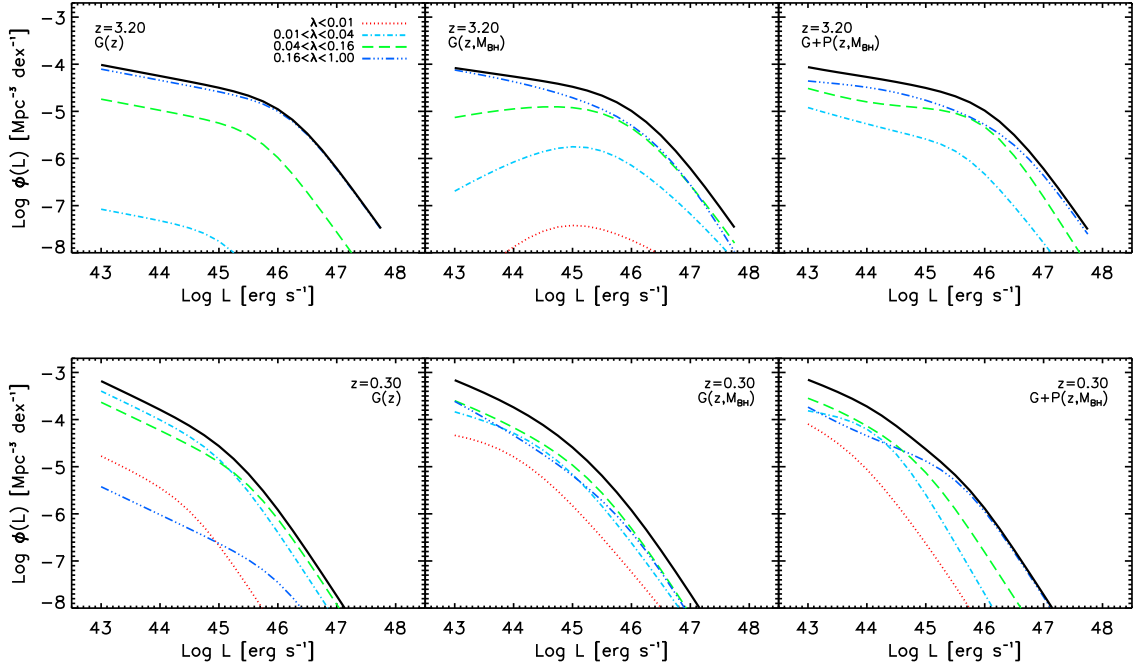}
    \caption{Contributions of different Eddington ratio distributions
    to the overall bolometric luminosity functions at $z=3.2$ (top) and $z=0.3$
    (bottom),
    as predicted by the \Gz, \GzM, and \GPzM\ models (left, middle, right
    panels, respectively).}
    \label{fig|FLlambdaContributions}
\end{figure*}

\begin{figure*}
    \includegraphics[width=15truecm]{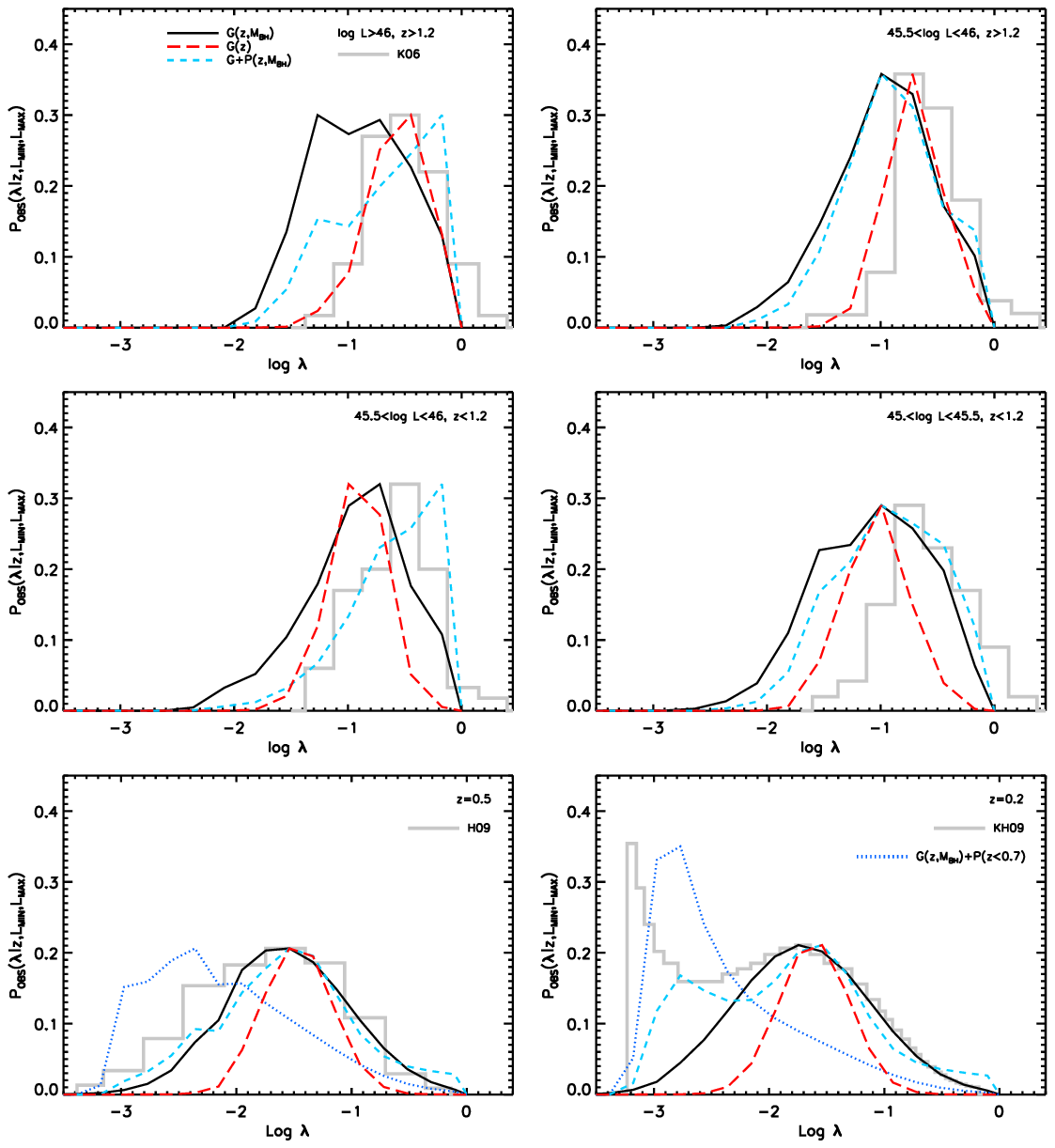}
    \caption{Similar format to Figures~\ref{fig|CompareKollmeier1} and
    \ref{fig|CompareKauff1}. Comparison among the \GzM, \Gz\ and
    \GPzM\ predicted Eddington ratio distributions (\emph{solid}, \emph{red
    long-dashed}, and \emph{cyan dashed} lines, respectively) compared to the
    K06 data at $z=2$ and $z=1$, respectively (\emph{top}
    and \emph{middle} panels), and to H09 at $z=0.5$
    (\emph{lower left}) and KH09 in the local
    Universe (\emph{lower right}).  In the lower panels, we also include
    curves for the steep power-law model introduced in
    Section~\ref{subsec|PLmassdependence2}.
    }
    \label{fig|CompareKollKauff}
\end{figure*}

In any model with a broad \PL,
the observable distribution of Eddington ratios depends on the
luminosity of the AGN being considered.  Fig.~\ref{fig|FLlambdaContributions}
shows the contribution to the bolometric luminosity function
from different ranges of $\lambda$ at $z=3.2$ (top panels) and $z=0.3$
(bottom panels) in the
\Gz, \GzM, and \GPzM$\,$ models (left, middle, and right panels, respectively).
There is a natural trend for the high end of the
luminosity function ($\log L > 46$) to be mainly contributed by high-$\lambda$
black holes, with a broader range of $\lambda$
contributing at lower luminosities.
Interestingly, we find that the $\lambda < 0.01$ range does not
dominate at any luminosity at any redshift.
Only in the \GzM\ model and at low redshifts are there roughly equal
contributions to the luminosity function from each
logarithmic bin of $\lambda$ in the range $0.01-1.0$.
This is important to compare with models that rely on
a significant contribution from ``ADAF-type'' modes
to the global accretion history of black holes
(see, e.g., Merloni \& Heinz 2008 and Draper \& Ballantyne 2010).

Fig.~\ref{fig|CompareKollKauff} compares the Eddington ratio
distributions of the \GzM, \Gz, and \GPzM\ models (black solid, red long-dashed,
and cyan dashed lines, respectively) to the observational estimates of K06,
H09, and KH09,
discussed earlier in Section~\ref{subsec|CompareWithObservedPL}.
Beginning at low redshift (bottom panels), we see that the
declining $\lambda_c(z)$ in the redshift-dependent models
resolves the discrepancy seen in Fig.~\ref{fig|CompareKauff1},
producing much better agreement with the H09
and KH09 data. The additional broadening in the Gaussian component
(Eq.~\ref{eq|Sigma})
produces excellent agreement with the H09 distribution
and excellent agreement with KH09 for $\lambda > 10^{-2.5}$.
The \GPzM\ model achieves similar agreement, matching KH09
slightly better at low luminosities but still falling short
at $\lambda < 10^{-3}$.
The blue dotted curves in the lower panels show the \GzM\ model with
the steep power-law at low $z$, introduced in
Section~\ref{subsec|PLmassdependence2}.  The predicted
Eddington ratio agrees poorly with both the H09 and KH09 data.

At $z>1.2$ (upper panels) model predictions are in approximate agreement
the K06 data, with \Gz\ showing the best agreement.
However, the $z=1$ model outputs tend to disagree with the K06, $z<1.2$
histograms (middle panels).
We note that Netzer \& Trakhtenbrot (2007) find a
peak at $\lambda \approx 0.1$ for $M=10^{8}-10^{8.5} M_\odot$
black holes at $z=0.7$. Kelly et al. (2010) have also recently claimed an Eddington ratio distribution
from SDSS of Broad Line Quasars that peaks at $L/L_{\rm Edd} \sim 0.05$
with a dispersion of $\sim 0.4$ dex.
Both these results would be closer to the model predictions.

While the duty cycle data considered in Section~\ref{subsec|PLmassdependence}
seem to favor a $\lambda_c$ that decreases at higher $\mbhe$,
the K06 data disfavor this solution.
We note that the luminosity threshold of the Xue et al. (2010)
study that motivates this model is much lower than that of the
K06 data, $L_{\rm min} \sim 10^{43}\ergse$ rather
than $L_{\rm min} \sim 10^{45}-10^{46}\ergse$.  A reconciliation
of these results could therefore lie in a model that behaves
differently in these two luminosity regimes.

\subsection{The Impact Of Mergers}
\label{subsec|Mergers}

\begin{figure*}
    \includegraphics[width=10truecm]{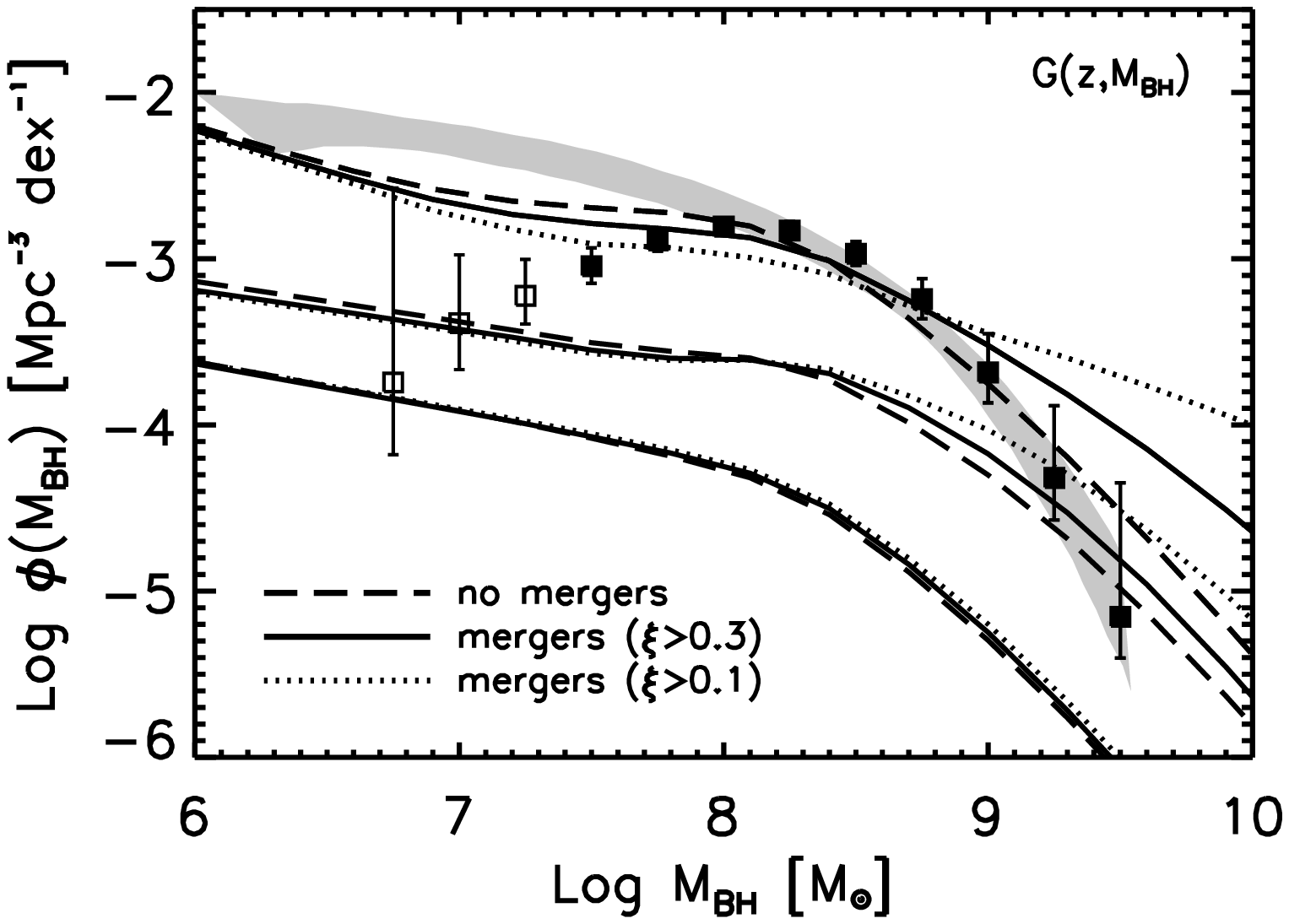}
    \caption{Predicted black hole mass function for the \GzM\ model
    without mergers (\emph{dashed} lines), with only major mergers
    (\emph{solid} lines),
    and with minor and major mergers (\emph{dotted} lines). The
    set of lines, from bottom to top, are the model predictions at $z=4, 2,
    0.1$, respectively. The data are as in Fig.~\ref{fig|BHMFandDuty}. The
    cumulative effect of mergers is minor at $z>2$ but becomes
    significant at low redshifts and high masses.}
    \label{fig|Mergers}
\end{figure*}

It is usually assumed that mergers of galaxies are followed by mergers
of their central black holes, making black hole mergers a potentially
important source of evolution in the black hole mass function
(e.g., Hughes \& Blandford 2003; Wyithe \& Loeb 2003;
Islam et al. 2004; Scannapieco \& Oh 2004; Yoo \& Miralda-Escud\'{e} 2004;
Volonteri et al. 2005; Lapi et al. 2006; Yoo et al. 2007; Marulli et al. 2008;
Bonoli et al. 2009; Shen 2009; SWM; Bonoli et al. 2010; Shankar 2010; Shankar et al. 2010a,d; Shankar et al. 2011;
Kocsis \& Sesana 2011; Kulkarni \& Loeb 2011).
We will save a full discussion of mergers for future work, but here
we briefly assess their potential impact on the mass function.
Improving on our simplified calculation in SWM, we here follow the schemes
proposed by Shen (2009) and Shankar et al. (2010b), computing the
rate of black hole mergers from the halo merger rate
predicted from fits to N-body dark matter simulations (Fakhouri \& Ma 2008),
corrected by a dynamical friction timescale.
In the presence of mergers
the continuity equation reads as
\begin{equation}
\frac{\partial n_{\rm BH}}{\partial
t}(M_{\rm BH},t)=-\frac{\partial (\langle \dot{M}_{\rm BH}\rangle
n_{\rm BH}(M_{\rm BH},t))}{\partial M_{\rm BH}}+ S_{\rm in} - S_{\rm out}\,,
\label{eqApp|contEqMergers}
\end{equation}
where $S_{\rm in}$ and $S_{\rm out}$
are, respectively, the merger rate of smaller mass black holes ending
up with mass \mbh\ and the merger rate of black holes with initial mass \mbh\
merging into more massive systems.
The merger rate of haloes as a function of redshift, mass ratio of the progenitors, and
mass of the remnants are taken from Fakhouri \& Ma (2008).
We then convert the merger rate of haloes to a merger rate of black holes
via the median \mbh-\mh\ relation,
defined, at all times, by cumulative number matching between
the black hole and halo mass functions,
which allows us to associate the proper halo
merger rate to a given bin of black hole mass.
Full details are given in
Appendices~\ref{Appendix|MbhMhaloRelation} and~\ref{Appendix|Mergers}.
We make the limiting case assumption that halo mergers are always
followed by black hole mergers after a dynamical friction time.
The accuracy of this assumption remains a matter
of debate (e.g., Cavaliere \& Vittorini 2000; Shen 2009; Shankar 2010,
and references therein).  The impact of the dynamical friction time delay
itself is irrelevant at $z \lesssim 2$ (see Shen 2009).

Fig.~\ref{fig|Mergers} shows the mass function of the \GzM\ model
at $z=4$, 2, and 0.1 without mergers (dashed lines), including major
mergers
above a black hole mass-ratio threshold $\xi=0.5$ (solid lines), and including
all mergers above a black hole mass-ratio threshold $\xi=0.1$ (dotted lines).
Mergers have limited effect at $z > 2$, but by $z=0.1$ they
have dramatically boosted the space density of black holes with
$\mbhe > 10^9 M_\odot$, to a level clearly inconsistent with the
SWM and Vika et al. (2009) estimates.
The discrepancy would be more severe if we did not include
mass-dependent $\epsilon$ in the \GzM\ model
(see Fig.~\ref{fig|BHMFcomparison}).
We caution, however, that
in this regime the local mass function estimates rely largely on
extrapolation of the $\mbhe-\sigma_*$ or $\mbhe-M_*$ correlations
into a range with limited observational constraints.  This is also
a mass range where typical galaxy hosts are gas poor, and it is
not clear that ``dry'' galaxy mergers necessarily lead to black
hole mergers.

In general, the inclusion of mergers makes it even harder
to fit the observed steep decline of black hole abundance at high masses.

\subsection{A $\lambda$-dependent bolometric correction}
\label{subsec|kbolLambda}

The inclusion of mergers puts otherwise acceptable models
at risk of overpredicting the high mass end of the local black
hole mass function.  We now discuss one possible resolution of
this tension, a $\lambda$-dependent bolometric correction.

\begin{figure}
    \includegraphics[width=8.5truecm]{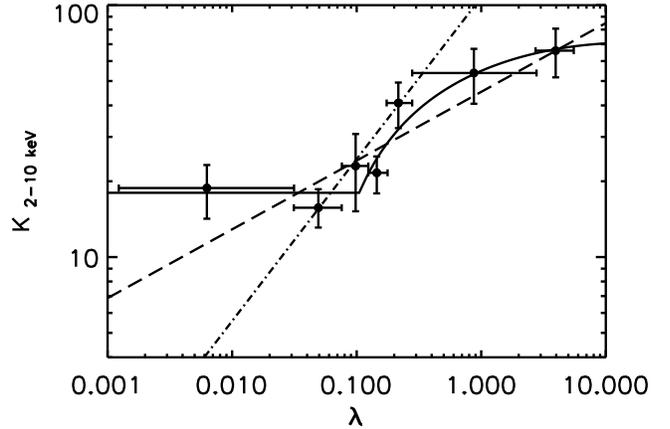}
    \caption{Binned data on the $2-10$ keV X-ray bolometric correction as
    a function of Eddington
ratio from Vasudevan \& Fabian (2007). The \emph{solid} line is the analytical
approximation used
in this paper, while the \emph{long-dashed} and \emph{dot-dashed} lines
are the results from Lusso et al. (2010).}
    \label{fig|KbolLambda}
\end{figure}

Following previous work
(e.g., Elvis et al. 1994; Marconi et al. 2004; Shankar et al. 2004;
Hopkins et al. 2007), we have so far assumed that the bolometric
correction depends only on bolometric luminosity (all the X-ray data
in SWM were converted to bolometric luminosities using the
$L$-dependent bolometric correction by Marconi et al. 2004). However, some
studies (e.g., Dai et al. 2004; Saez et al. 2008; and references
therein) show signs for variations in the spectral energy
distributions of AGNs with either redshift or accretion properties.
In particular, Vasudevan \& Fabian (2007, 2009) suggest that variations in
the disc emission in the ultraviolet may be important to build the
optical-to-X-ray spectral energy distributions of AGNs. From a
sample of 54 AGNs from the Far Ultraviolet Spectroscopic Explorer
(FUSE) and X-ray data from the literature, they claim evidence
for a large spread in the bolometric corrections, with no
simple dependence on luminosity being evident. Their results suggest
instead a more well-defined relationship between the bolometric correction
and Eddington ratio, with a transitional region at an Eddington
ratio of $\sim 0.1$, below which the X-ray bolometric correction is
typically 15-25, and above which it is typically 40-70. As shown in
Fig.~\ref{fig|KbolLambda}, we approximate their results by setting
\begin{equation}
K_{\rm 2-10 keV}(\lambda)= \left\{
  \begin{array}{lll}
    18 & \hbox{if $\lambda \le 0.105$} \\
    54.85+26.78\, \log L_X-\\
    -11.11 \, (\log
L_X)^2 & \hbox{if $0.105\le \lambda \le 1$\,\, .}
  \end{array}
\right. \label{eq|BClambda}
\end{equation}
Our analytic approximation is shown as a solid line in
Fig.~\ref{fig|KbolLambda},
against the binned data by Vasudevan and Fabian (2007; the fit and the data
extend to $\lambda>1$), while the
dot-dashed and long-dashed lines are more recent fits from Lusso et al. (2010)
derived
from a larger sample. As the latter authors point out, while a trend
of bolometric correction with \lam\ might exist, determining its exact
slope is still challenging given the large dispersion in the data. In this
section we will use Eq.~\ref{eq|BClambda} as a reference, noting that
the Lusso et al. (2010) fits
provide consistent results in the range of interest here.

\begin{figure}
\includegraphics[width=8.5truecm]{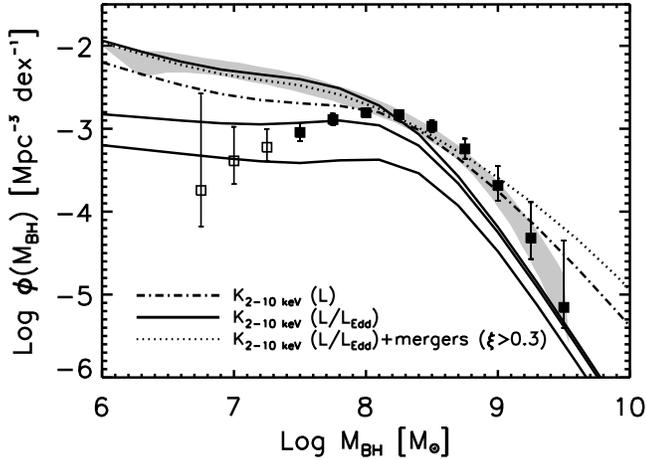}
\caption{Influence of a $\lambda$-dependent bolometric correction
on the predicted black hole mass function. The {\emph dot-dashed} line
shows the $z=0$ mass function of the
\GzM\ model with our standard bolometric correction.
The {\emph solid} lines show the evolving mass function at $z=2$, 1, and 0
for the same model assuming the $\lambda$-dependent bolometric
correction of Eq.~\ref{eq|BClambda}. The dotted line is the $\lambda$-dependent \GzM\ model inclusive of mergers (with $\xi >0.3$).  Data points and grey band are the same
as those in Figs.~\ref{fig|BHMFandDuty} and~\ref{fig|BHMFcomparison}.}
\label{fig|NMKbolLambda}
\end{figure}

In our numerical formalism described in
Appendix~\ref{app|solvingContEq}, it is
straightforward to insert a \lam-dependent bolometric correction. Given that
the relation $L\propto K_{\rm 2-10 keV}(\lambda) \times L_X \propto \lambda
\mbhe$, it implies that $L_X\propto [\lambda/K_{\rm 2-10 keV}(\lambda)] \mbhe
$. Therefore, having a \lam-dependent bolometric correction is equivalent
to running the code replacing bolometric luminosities with
$L_X$. We thus solve Eq.~\ref{eq|PhiLLambda} for computing the duty cycle
by replacing the bolometric luminosity function
$\Phi(L,z)$ on the left-hand side with the X-ray luminosity function $\Phi_X(L_X,z)$,
and using effective Eddington ratios $\lambda'=\lambda/K_{\rm 2-10 keV}(\lambda)$.

Most predictions of our models are not sensitive to this change
of bolometric correction. In particular, we have
checked that after taking into account the $\lambda$-dependent
conversions between sample flux limits and bolometric
luminosities, both duty cycles and Eddington
ratio distributions showed similar behaviours
to the ones predicted by the \GzM\ model with a luminosity-dependent
bolometric correction.

Nevertheless, the $\lambda$-dependent
correction does have a significant effect on the low-$z$ black
hole mass function.
The solid lines (at $z=0,1,2$, from top to bottom)
in Fig.~\ref{fig|NMKbolLambda} show the \GzM\ model with a $\lambda$-dependent bolometric correction as in Eq.~\ref{eq|BClambda}.
High-mass black holes have preferentially lower $\lambda$ in this model,
and the lower bolometric luminosity at a given X-ray luminosity
reduces the inferred growth rate of these massive black holes.
Since \GzM previously agreed well with the $z=0$ mass function,
it now underpredicts the high mass end.

A $\lambda$-dependent bolometric correction would reduce the need
for mass-dependent $\epsilon$ in our \GzM\ and \GPzM\ models
(Section~\ref{subsec|epsilonmass}, Fig.~\ref{fig|BHMFcomparison}), which
was inferred partly from the high mass end of the local mass
function, though it was also supported by
direct empirical evidence for mass-dependent $\epsilon$.
Alternatively, a $\lambda$-dependent
bolometric correction could compensate the impact of mergers on our
standard \GzM\ model (see Fig.~\ref{fig|Mergers}) to yield improved
agreement with local mass function estimates.  The dotted line
in Fig.~\ref{fig|NMKbolLambda} is the predicted $z=0$ black hole mass
function for the \GzM\ model including a mass-dependent $\epsilon$,
$\lambda$-dependent bolometric correction, and black hole mergers
with $\xi > 0.3$.  Agreement with observational estimates is
significantly improved relative to the standard bolometric
correction case (uppermost solid line in Fig.~\ref{fig|BHMFcomparison}).
Uncertainties in bolometric corrections --- their normalization
and their dependence on luminosity, $\lambda$, or other factors --- remain an important source of uncertainty when
testing evolutionary models of the black hole population
against the local census of black holes.

\subsection{Specific Black Hole Accretion Rate}
\label{subsec|SBHAR}

\begin{figure*}
    \includegraphics[width=15truecm]{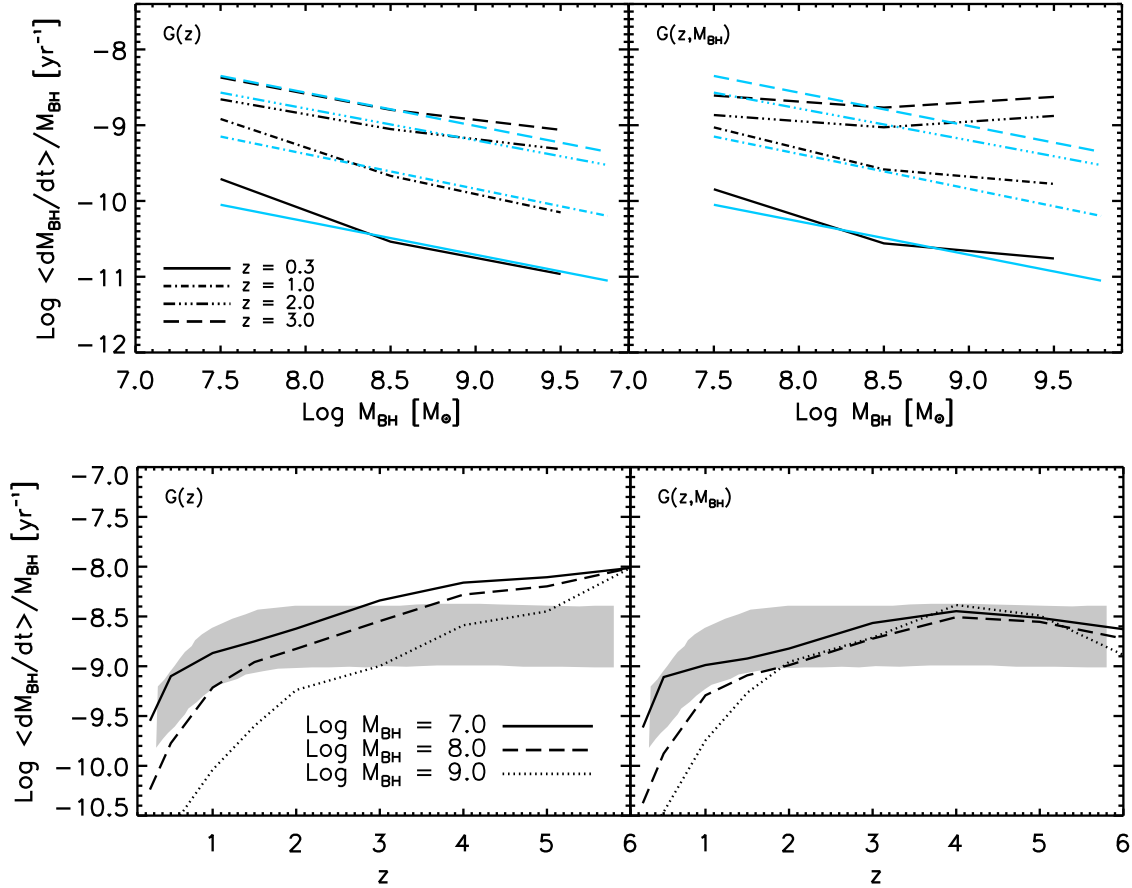}
    \caption{Mean specific black hole accretion rate
    (Eq.~\ref{eq|SBHAR})
    predicted by
    the \Gz\ (\emph{left}) and \GzM\ (\emph{right}) models as
    a function of black hole mass (\emph{top}) and of redshift
    (\emph{bottom}).  In the top panels,
    \emph{cyan} lines are the mean specific star formation rates as calibrated
    by Karim et al. (2011; their Table 4), with stellar masses simply scaled
    to black hole masses assuming a proportionality factor of $10^{-3}$, as
    measured in the local Universe. In the bottom panels,
    the \emph{grey area} is the specific star
    formation rate with error bars calibrated by Gonz{\'a}lez et al. (2010)
    for galaxies with stellar mass $\sim 5\times 10^9 - 10^{10} \msune$,
    which can be compared to the predictions for
    $10^7\msune$ black holes (\emph{solid} curves).}
    \label{fig|SBHAR}
\end{figure*}

Several authors have noted
that the average black hole accretion rate
has a redshift dependence morphologically similar
to the cosmological SFR (e.g., Marconi et al. 2004; Merloni et al. 2004;
Silverman et al. 2008a; Zheng et al. 2009; SWM).
Fig.~\ref{fig|SBHAR} adds a new piece of information to
the co-evolution of black holes and galaxies, plotting the mean specific
accretion rate of black holes of a given mass at a given redshift
defined as
\begin{equation}
\label{eq|SBHAR}
\frac{\langle \dot{M}(\mbhe,z) \rangle}{\mbhe} = \frac{\langle \lambda(\mbhe,z)
\rangle U(\mbhe,z)}{t_s} \, .
\end{equation}
While the mean accretion rate
does not depend on the \PL\ distribution
but only on the assumed radiative efficiency,
the mean specific accretion rate depends significantly on the input \PL,
so it provides diagnostic power beyond that in the global rates.

In the \Gz\ model (left panels), the specific mean accretion rate declines with
increasing black hole mass at all epochs.
Cyan lines in the upper panels show the mean specific star
formation rate (SSFR) recently derived by Karim et al. (2011; see also Noeske
et al. 2009) at the
same redshifts (from their Table 4),
with stellar masses simply re-scaled by a factor of $10^{-3}$ to convert to
black hole masses.
In this simple comparison with the \Gz\ model,
the mean specific black hole accretion
rate decreases with mass and increases with redshift in a remarkably
similar way as the mean star formation rate, with a slope
$\sim \mbhe^{-0.4}$.
However, when we consider the \GzM\ model (right panels), which better matches
data on black hole duty cycles and $\lambda$-distributions,
we predict higher specific accretion rates at high masses,
reflecting the higher duty cycles in this model.
For simplicity, here we use a constant radiative efficiency,
thus a constant $t_s$ in Eq.~\ref{eq|SBHAR}.
Silverman et al. (2009) also find that in relatively massive
high-$z$ galaxies the ratio between average black hole accretion rate and SFR
is higher by up to an order of magnitude with respect to the classical
$10^{-3}$, in fair agreement with the prediction of the \GzM\ model,
though dependent on the actual variations of radiative efficiency with mass and/or time. In fact, we checked that including a mass-dependent radiative efficiency as in Eq.~\ref{eq|radefficiencyMass} would line up the specific black hole accretion rate with
the SSFR at all masses.

The lower panels of Fig.~\ref{fig|SBHAR}
show the mean specific black hole accretion rate
as a function of redshift for different bins of black hole mass, as labelled.
The grey bands indicate the uncertainties around the measured SSFR as
catalogued and derived by Gonz{\'a}lez et al. (2010) for galaxies with
stellar mass $\sim 5\times 10^9 - 10^{10} \msune$.
The latter should be compared with only the specific accretion rate onto
black holes with current mass $\sim 10^7 \msune$ (solid lines), but for
completeness
we also show the accretion rate for more massive black holes
$\mbhe = 10^8 \msune$ and $\mbhe =  10^9 \msune$ (long-dashed and dotted
lines, respectively), as labelled.
Overall, consistently with what is found in the upper panels,
the models predict an accretion rate that tracks the SSFR,
though the \Gz\ model tends to produce a specific accretion rate that
steadily increases even at $z \gtrsim 2$, at variance with the data for
the star formation rate.
The \GzM\ model predicts a redshift dependence of the accretion rate
morphologically similar
to that of galaxies, a fact that might play some role
in explaining the still puzzling plateau at $z \gtrsim 2$ of the galactic
SSFR$(z)$, which is poorly reproduced by semi-analytic models and might
require some extra source of early feedback (e.g., Weinmann et al. 2011).

\section{Discussion}\label{sec|discu}

Although we have considered a wide variety of models, with different
\PL\ shapes and different redshift and mass dependences of $\lambda_c$
and $\epsilon$, every one of these models shows significant (factor
of several) disagreement
with at least one of the observational tests we have examined.
This failure could indicate that our model assumptions are still
too restrictive to describe the real black hole population --- for
example, we generally assume that the shape of \PL\ is independent of redshift
and black hole mass except for overall shifts in $\lambda_c$, and
we have considered restricted functional forms for these $\lambda_c$ trends
and for \PL\ itself.  Alternatively, the problem could lie in one
or more of the data sets themselves, since these are frequently
derived from noisy or uncertain estimators (e.g., for black hole
masses) and from input samples that are subject to selection
biases and incompleteness (e.g., for active galaxy fractions).
Here we highlight aspects of the data that appear especially
difficult to reproduce within our class of models, or where different
data sets appear to drive the models in contradictory directions.
Recall that all of our models reproduce SWM's estimate of the
bolometric luminosity function (summarized in Section~\ref{subsec|AGNLF})
by construction.  SWM discuss remaining uncertainties in this
luminosity function and their impact on inferred model parameters.

The first tension within the data is at the low-mass end of the local
black hole mass function (see Fig.~\ref{fig|BHMFandDuty}), where SWM
(and a number of other studies) find a $\Phi(\mbhe)$ rising
to low masses but Vika et al. (2009) and some other studies
(e.g., Graham et al. 2007) find a falling $\Phi(\mbhe)$.  This region of
the mass
function remains difficult to probe because of uncertainties
in bulge-disk decomposition and because
the black hole mass correlations for spiral bulges
are more uncertain than those for high mass, bulge-dominated galaxies.
In our models, it is difficult to produce a turnover like that
of Vika et al. (2009), though the \Gz\ model goes in this direction
(Fig.~\ref{fig|BHMFcomparison}) because it ascribes much of the low
luminosity AGN activity to low-$\lambda$ accretion by massive
black holes, hence reducing the growth of low mass black holes.
However, the predicted mass function in this regime depends on the
input luminosity function in a range that is largely extrapolated
from brighter magnitudes.  Therefore if $\Phi(\mbhe)$ really does
turn over at low masses, a plausible explanation is that the faint
end of the AGN luminosity function is flatter than the one adopted here.

A second tension, with more serious implications for the issues
at the core of this paper, is between the narrow Eddington ratio
distributions measured by K06 and the broader
distribution, peaking at lower $\lambda_c$, measured by H09.
The two data sets overlap in redshift, though H09
are at the low redshift end of the K06 range.
H09, and KH09 at still lower $z$, favor a
\PL\ that is broad in shape (like G+P) and evolving to low $\lambda_c$
at low redshift, but it is difficult to reconcile such a model with
the K06 histograms.  The luminosity thresholds
for the H09 and K06 data sets are very different, roughly
$L_{\rm min} = 10^{43}\ergse$ and $L_{\rm min}=10^{45}\ergse$,
respectively, so there is no direct contradiction between
the measurements. Possibly a model that allows a different
{\it shape} of \PL\ for high and low mass black holes, or that
allows a power-law that is steeper or further offset from the
log-normal peak, could be made consistent with both sets of
observations. The two samples might differ
for more profound physical reasons, as recently emphasized by
Trump et al. (2011), who showed that broad lines
might disappear at $\lambda < 0.01$ because of a change in
accretion flow structure.

The observed \PL\ distributions also include errors
in black hole mass estimations, so the intrinsic distributions
should in principle be even narrower.
Correcting for this observational broadening would exacerbate
the tension with our \GzM\ and \GPzM\ models, which already
predict broader \PL\ distributions than those found by K06.

Another important tension arises between the low duty cycles found
for low redshift quasars by Schulze \& Wisotzki (2010) and the much
higher active galaxy fractions found by Kauffmann et al. (2003),
with a gap that is roughly two orders of magnitude.  The difference
between Type 1 and Type 2 AGN could possibly explain a factor of $10-20$
(Greene \& Ho 2009), though this factor is already
large compared to conventional estimates of obscured-to-unobscured
AGN ratios, and other empirical and physical effects might
need to be invoked to explain
such a strong discrepancy (e.g., Trump et al. 2011).
By adopting the Greene \& Ho (2009) ratios in our predictions, we
find models that are roughly consistent with
both Kauffmann et al.\ (2003) and
Schulze \& Wisotzki (2010), but even our best cases disagree
with one of these data sets by a factor of several in some
black hole mass range (see Fig.~\ref{fig|DutyCyclez0}).
We have converted Kauffmann et al.'s [OIII] luminosities
to bolometric luminosities assuming a constant bolometric correction,
and scatter or biases in this correction
(see, e.g., Capetti\ 2011) might account for some of the
discrepancy with Schulze \& Wisotzki (2010).
Best et al.\ (2005, see their Fig.~2) also find high AGN
fractions for SDSS galaxies --- $20-40\%$ for
$L_{\rm [OIII]} > 10^{5.5}L_\odot$ ($\log L \ga 42$) --- comparable
to those of Goulding et al.\ (2010).

A fourth tension arises from the trend of higher active fractions
for more massive galaxies found by Xue et al. (2010) and
Bundy et al. (2010).  If \PL\ does not evolve, then matching
observed AGN luminosity evolution leads to downsizing, i.e.,
a {\it decrease} of duty cycle with increasing black hole mass
(Fig.~\ref{fig|BHMFandDuty}).  A $\lambda_c(z)$ that declines
towards low redshift can soften this trend, but within
our considered range of models, it does not eliminate downsizing
entirely. We have been able to produce a trend of rising duty
cycle with rising black hole mass by making $\lambda_c$ decrease
towards higher \mbh, but the resulting models then tend to be inconsistent
with the K06 Eddington ratio distributions, which
do not show such a trend (Fig.~\ref{fig|CompareKollKauff}).
Furthermore, the model trends of $U(M_{\rm BH})$ remain
flatter than those found by Xue et al. (2010) and Bundy et al. (2008),
and these models still yield falling $U(M_{\rm BH})$ at $z=0.5$.
One caveat is that we are translating the observed galaxy stellar
masses to corresponding black hole masses assuming a linear relation
with no scatter.  If the scatter between black
hole mass and galaxy mass were large at these redshifts, then
the trend between active fraction and galaxy mass could
be partly induced by a higher probability for black holes of fixed mass to
be active if they reside in more massive galaxies.

In Section~\ref{subsec|PLmassdependence2} we considered a model
in which \PL\ rises steeply towards low $\lambda$
(with $P\propto \lambda^{-0.9}$), which helps produce a rising $U(M_{\rm BH})$
trend for a luminosity thresholded sample because more massive black
holes can radiate at low $\lambda$ while remaining above
threshold (Aird et al. 2011; Mainieri et al. 2011).
We restricted this steep \PL\ to $z<0.7$, since at higher redshifts
it leads to duty cycles above unity.
With this model (inspired by that of Aird et al.\ 2011), we are able
to obtain a rising $U(M_{\rm BH})$ at $z=0.5$.  However, the prediction
for the observed \PL\ disagrees with the data of K06, H09, and KH09,
a discrepancy also noted by Aird et al.\ (2011).

A final tension arises at the high end of the black hole mass
function, where the models tend to overpredict the data.
While models with a mass-independent $P(\lambda)$ are acceptable
within the estimated
observational uncertainties, adding mass dependence of
$\lambda_c$ to produce a rising $U(\mbhe)$ leads to a
substantial overprediction of the abundance of the most massive black
holes. We have mitigated this problem in
our \GzM\ model by also increasing the radiative efficiency
at high \mbh\ (Fig.~\ref{fig|BHMFcomparison}).
This solution has some observational support, as discussed
in Section~\ref{subsec|PLmassdependence}. On the other hand, including mergers as
calculated in Section~\ref{subsec|Mergers} worsens the overprediction
of $\Phi(\mbhe)$ at high masses (Fig.~\ref{fig|Mergers}).

  There are other factors that may ameliorate the discrepancy between
models and observational determinations of the abundance of the most
massive black holes.
A $\lambda$-dependent (or simply lower) bolometric correction may resolve this
tension, as discussed in Section~\ref{subsec|kbolLambda}.
The model shown by the dotted curve in Fig.~\ref{fig|NMKbolLambda},
which includes mergers, is acceptable within current uncertainties.
The high end of the mass function relies on
extrapolation of the \mbh-bulge relations into a regime where
there are few calibrating galaxies
(Schulze \& Gebhardt 2011).
Another effect that can help fitting the AGN luminosity function
with a steeper decline of the black hole mass function at high masses
is anisotropic emission by AGNs, which would give rise to apparent
values of $\lambda$ (i.e., inferred from the observed flux in our
direction under the assumption of isotropy) that are occasionally larger
than unity. The majority of the most luminous AGN would then correspond
to objects that have their brightest direction of emission pointing to
us, rather than to AGN with the highest black hole masses,
and the mass accretion rates of these AGN would be lower than
their observed luminosities suggest.
(Note that in
our models we have imposed a cutoff on the $P(\lambda)$ distribution at
$\lambda > 1$.)

If, in light of this discussion, we take a rather generous
view of the systematic uncertainties in the observational constraints
we have considered, then our \GzM\ and \GPzM\ reference models may
be viewed as at least moderately successful, while the other
reference models --- $\delta$, G, G+P, and \Gz\ --- all fail
drastically on at least one observable.  Table~\ref{table|models2}
summarizes the observational comparison based on an admittedly
subjective assessment of the results shown in
Figures~\ref{fig|BHMFandDuty}, \ref{fig|CompareKollmeier1}, \ref{fig|CompareKauff1}, \ref{fig|DutyCyclez0}, \ref{fig|BHMFcomparison}, and \ref{fig|CompareKollKauff}.  We consider as constraints
the SWM estimate of the local black hole mass function, the
high-$z$ Eddington ratio distributions from K06, the low-$z$
Eddington ratio distributions from H09 and KH09, and the
duty cycle estimates from active galaxy fractions at $z\approx 0.2$
shown in Figure~\ref{fig|DutyCyclez0}.  Recall that all models reproduce our input
AGN luminosity function by construction.  We assign a
$\checkmark$ when a model reasonably describes an observation
with some allowance for systematic uncertainty, an X when it
clearly fails, and a $-$ for intermediate cases.
The non-evolving models and \Gz\ model all fail to match
the low-$z$ \PL\ or the low-$z$ $U(\mbhe,L_{\rm min})$,
or both.  The \GzM\ and \GPzM\ models have no such drastic
failures, though they are far from perfect matches to the data.
However, none of our models reproduce the trend of duty cycle
with black hole mass illustrated in Figure~\ref{fig|DutyCycleHighz}, though the
\GzM\ model is the least discrepant.

\begin{table*}
\begin{tabular}{|l|l|l|l|l}
  \hline
  Model & \Phibh $, \, z=0$ & \PLMz $, \, z>0.5$ & \PLMz $, \, z\le 0.5$ & $U(\mbhe,\Lmine,z)$ \\
  \hline
  \hline
  $\delta$ & $\checkmark$ & X & X & X \\
  G & $\checkmark$ & $\checkmark$ & X & X \\
  G+P & $\checkmark$ & $\checkmark$ & X & X \\
  G(z) & -- & $\checkmark$ & X & $\checkmark$ \\
  G(z,$\mbhe$) & $\checkmark$ & -- & -- & $\checkmark$ \\
  G+P(z,$\mbhe$) & $\checkmark$ & $\checkmark$ & $\checkmark$ & -- \\
  \hline
  \end{tabular}
  \caption{List of the Reference Models listed in Table~\ref{table|models} along
   with a qualitative assessment of their agreement with the data. We assign a
$\checkmark$ when a model reasonably describes an observation
with some allowance for systematic uncertainty, an X when it
clearly fails, and a $-$ for intermediate cases. The \PLMz\ at $z>0.5$ column refers to the K06 data (Figures~\ref{fig|CompareKollmeier1} and \ref{fig|CompareKollKauff}). The \PLMz\ at $z\le 0.5$ refers instead to the H09 and KH09 data (Figures~\ref{fig|CompareKauff1} and
\ref{fig|CompareKollKauff}). Finally, the $U(\mbhe,\Lmine,z)$ column refers to the multiple data sets reported in Figures~\ref{fig|DutyCyclez0} and \ref{fig|DutyCycleHighz}.}
  \label{table|models2}
\end{table*}

\section{CONCLUSIONS}
\label{sec|conclu}

We have extended the formalism of continuity-equation modeling
of the black hole and AGN populations to allow a distribution
of Eddington ratios \PL.  With this broader class of models
we have addressed two new categories of observations, direct
estimates of Eddington ratio distributions of active black holes
and estimates of duty cycles from active galaxy fractions.
Both of these categories have been areas of intense observational
investigation over the last five years, and as representative
examples we have concentrated on \PL\ estimates from
Kollmeier et al.\ (2006; K06) at $z \geq 1$,
Hickox et al.\ (2009; H09) at $z \approx 0.5$, and
Kauffmann \& Heckman (2009; KH09) at $z \approx 1$, and on
active galaxy fraction data from Bundy et al.\ (2008) and
Xue et al.\ (2010) at $z \geq 0.5$ and from
Goulding et al.\ (2009), Kauffmann et al.\ (2003), and
Schulze \& Wisotzki (2010) at low redshift.
We account for the effective luminosity thresholds of these
analyses in our model predictions, though the thresholds are
not always clearly defined.

As forms for \PL\ we consider a Gaussian in $\log\lambda$ (G)
and a Gaussian with a power-law extension to low-$\lambda$ (G+P).
If the radiative efficiency $\epsilon$ and characteristic
Eddington ratio $\lambda_c$ (where the Gaussian peaks) are
held fixed, then changing from a single $\lambda$ value to either
of these distributions has little impact on the evolution of
the black hole mass function inferred from continuity-equation modeling.
As in the single-$\lambda$ models of SWM (and similar models by
Shankar et al.\ 2004 and Marconi et al.\ 2004), we find that models
incorporating the observed AGN luminosity function and parameter
values $\epsilon \approx 0.07$ and $\lambda_c \approx 0.25$
yield a good match to observational estimates of the black hole
mass function in the local Universe.
The predicted black hole duty cycle declines rapidly with
decreasing redshift at $z<2$, and at redshifts $z \leq 1$
it declines sharply with increasing black hole mass over
the range $10^7 M_\odot - 10^9 M_\odot$
(``downsizing'' evolution).
An unattractive feature of the G+P model is that it requires
a high space density of massive black holes already present
at $z=6$, since only the small fraction of black holes with
high $\lambda$ are luminous enough to contribute to the
observed luminosity range at this redshift.  This large ``seed''
population appears physically unrealistic, and the low implied
duty cycle for high-luminosity quasars contradicts evidence from
their strong observed clustering at $z \approx 4$
(Shen et al.\ 2007; White et al.\ 2008; Shankar et al.\ 2010b).
We conclude (in agreement with Cao 2010) that any extended low-$\lambda$
tail of \PL\ must develop at lower redshifts ($z \leq 3$) rather than
being a redshift-independent feature of black hole fueling processes.

Our models with redshift-independent \PL\ predict duty cycles at
$z\leq 1$ that are far below observational estimates from active
galaxy fractions, and they do not match the low-$z$ Eddington
ratio distributions estimated by H09 and KH09.
Motivated by these discrepancies, we introduce the \Gz\ model with
a redshift-dependent $\lambda_c$ (Eq.~\ref{eq|PLz}), shifting
\PL\ to lower Eddington ratios at low redshift.
This model yields better agreement with the observations.
The low-redshift \PL\ remains narrow compared to the H09 and KH09
estimates, and the model still predicts a duty cycle
that declines with increasing black hole mass, in contradiction
to the Bundy et al.\ (2008) and Xue et al.\ (2010) finding
of higher AGN fractions in more massive galaxies.
We therefore introduce an additional {\it mass} dependence of
$\lambda_c$ (Eq.~\ref{eq|PLmassDependent};
$\lambda_c \propto M_{\rm BH}^{-0.3}$) at each redshift,
which, relative to mass-independent models, maps massive black
holes to lower luminosity, more numerous AGN.  At the same time,
we introduce a steady broadening of the Gaussian \PL\ towards
low redshift (Eq.~\ref{eq|Sigma}), or, for the G+P model,
a steady drop in the minimum Eddington ratio (Eq.~\ref{eq|LambdaMin})
that removes the need for an unrealistic seed population at $z=6$.

The predictions of the resulting models, \GzM\ and \GPzM,
are fairly similar.  Both models achieve a reasonable match to
the H09 and KH09 Eddington ratio distributions, though they
underpredict the KH09 distribution at $\lambda \leq 10^{-2.5}$.
They produce approximate agreement with the low-redshift duty
cycle estimates if we adopt the large (factors of $10-20$) ratios
of Type II to Type I AGN advocated by Greene \& Ho (2009),
though there are still factor of several discrepancies for
some data sets at some $M_{\rm BH}$ values.
Both models predict duty cycles that are flat or weakly
rising with black hole mass at $z>1$, thus improving the
agreement with Bundy et al.\ (2008) and Xue et al.\ (2010),
though neither model can reproduce the Xue et al.\ (2010)
trend at $z=0.5$.

Our models with mass-dependent \PLzM\ exhibit tension
with the K06 Eddington ratio distributions,
especially at $z\approx 1$, since the redshift and mass
dependence of $\lambda_c$ drive the predicted \PL\ distributions
to peak at $\log\lambda \approx -1$ while the
observed distributions peak at $\log\lambda \approx -0.6$.
Adding a steep ($P \propto \lambda^{-0.9}$) power-law at late times,
similar to the model of Aird et al.\ (2011), can better
match the Xue et al.\ (2010) duty cycle data at $z=0.5$, but it also spoils
the match with the H09 and KH09 Eddington ratio distributions.

Boosting the duty cycle, and thus the growth, of massive
black holes tends to overproduce the high-mass end of the
local mass function relative to observational estimates.
Our standard versions of the \GzM\ and \GPzM\ models therefore
incorporate a mass-dependent radiative efficiency
(Eq.~\ref{eq|radefficiencyMass}), following Cao \& Li (2008).
The higher efficiency assumed for higher mass black holes
reduces their inferred growth, restoring agreement with the
local mass function.  Alternatively, a $\lambda$-dependent
bolometric correction (Vasudevan \& Fabian 2007) can lower the
inferred growth of massive (hence lower $\lambda$) black holes
in these models, obviating the need for mass-dependent $\epsilon$.
However, black hole mergers at the rate suggested by black hole
merger statistics also raise the high mass end of the mass
function at low redshifts, so in a complete model the mass-dependent
$\epsilon$ may be be required even with the $\lambda$-dependent
bolometric correction.  Furthermore, Davis \& Laor (2011;
but see also Raimundo et al. 2011 and Laor \& Davis 2011) have
presented direct evidence for mass-dependent $\epsilon$ from
quasar spectral energy distributions, and such a dependence
could also explain the discrepancy between the $\epsilon \ga 0.2$
inferred for luminous, strongly clustered quasars at $z=4$
(Shankar et al.\ 2010b) and the average $\epsilon \approx 0.07$
implied by matching the local black hole mass density
(SWM and numerous references therein).  Each of these arguments
for mass-dependent $\epsilon$ rests on somewhat shaky ground, but
they all point in the same direction.
At $z>1$, the black hole mass functions implied by our models
are fairly insensitive to mergers and only moderately sensitive
to other model assumptions.

In agreement with previous studies, we find that the growth of the
black hole population tracks the overall evolution of the cosmic
star formation rate.  We also find generally good agreement
between the mean {\it specific} black hole growth rates
$\langle \dot{M}_{\rm BH}\rangle / M_{\rm BH}$ and
Karim et al.'s (2011) estimates of the specific star-formation
rates of star-forming galaxies over the full range $0.3 \leq z \leq 3$,
if we simply translate stellar masses to black hole masses
with a constant scaling factor of $10^{-3}$.

Over the last five years, measurements of AGN clustering have
improved dramatically in precision, redshift extent, and luminosity range.
These measurements provide valuable constraints on the relation between
active black holes and their host dark matter halos, which we have
previously explored in the context of single-$\lambda$ accretion
models (Shankar et al.\ 2010b,c).  We will extend our clustering
studies to include \PL\ distributions and mergers in future work.

With \PL\ and radiative efficiency allowed to depend on redshift
and black hole mass, our models have become rather elaborate
despite their simple physical basis. This complexity reflects
the growing richness of the observational data, especially the
measurements of Eddington ratio distributions and active galaxy
fractions over a wide range of redshift, mass, and luminosity,
from a variety of data sets. In order of decreasing robustness,
our key qualitative conclusions are
(a) that the characteristic Eddington ratio $\lambda_c$ declines
at low redshift, (b) that the \PL\ distribution broadens
at low redshift, (c) that more massive black holes have lower $\lambda_c$,
and (d) that more massive black holes have higher radiative efficiency.

Despite the flexibility of our
models, and despite investigating many variants beyond those
discussed in the paper, we have not found a model that fully reproduces
all of the observational constraints we have considered.
The remaining discrepancies presumably reflect some combination
of inadequate models and systematic errors in the data
sets, as discussed in detail in Section~\ref{sec|discu}.
On the model side, we have assumed restricted functional
forms for the mass and redshift dependence of \PL, and more
general behavior or sharper evolutionary transitions may
be required to match the data.  Observationally, estimates
of black hole masses and bolometric luminosities are both
subject to systematic uncertainties, and even when these
estimates are correct in the mean, scatter can have important
effects on inferred trends and distributions.  Many of the tensions
between the models and the data and among the data sets
themselves revolve around the seemingly disparate trends found
for optically luminous, broad-line quasars and varieties
of low luminosity, Type II AGN.  We have followed standard
practice in relating these two populations by a simple
obscuration factor, but a more nuanced relation between the
different categories of active black holes (e.g., Trump et al.\ 2011)
may be crucial to resolving some of the tensions highlighted here.

Continuity-equation models draw on the inevitable link between luminosity
and mass accretion to tie the observable population of AGN to the
evolving population of supermassive black holes that power them.
They provide a powerful framework for linking empirical studies that
probe a variety of observables across a wide span of redshift,
luminosity, and black hole mass.  As these empirical studies continue
to improve in precision, dynamic range, and control of systematic
uncertainties, they will refine the models
into a tightly constrained history of the cosmic black hole population.


\section*{acknowledgments}
FS acknowledges support from the Alexander von Humboldt Foundation and a Marie Curie Grant.
We also acknowledge support from NASA Grant NNG05GH77G and NSF grant
AST-1009505. DW acknowledges support of an AMIAS membership at the
Institute for Advanced Study during part of this work.
We thank Elisa Binotto, Adam Steed, Jaiyul Yoo,
Yue Shen, Chris Onken, Zheng Zheng, Jeremy Tinker, Pavel Denisenkov, Roderik Overzier,
Zoltan Haiman, Massimo Dotti, Peter Behroozi, Lucia Ballo, Ana Babi\'{c}, Lance Miller,
Vincenzo Mainieri for many interesting and helpful discussions.
We thank the anonymous referee for a constructive report
that helped us improve the clarity of the paper.

{}


\appendix

\section{Solving the continuity equation with broad input Eddington ratio
distributions}
\label{app|solvingContEq}

There is not a unique way to solve Eq.~\ref{eq|PhiLLambda}. Steed \& Weinberg
(2003)
adopted an input parametrical double power-law duty cycle,
and tuned the parameters to predict the AGN luminosity function.
More recently, a similar technique has been
adopted by Cao (2010), who performed a detailed
$\chi^2$ to up-to-date AGN luminosity functions
at all redshifts, and then solved the continuity
equation having full information on the input
duty cycle.

Following these previous works, and in particular Cao (2010),
at any redshift $z$ we parameterize the input active mass
function \Nact, i.e., the product of the duty cycle and the
black hole mass function, by a double-power law of the type
\begin{equation}
\Nacte=\frac{N_0}{\left( \frac{\mbhe}{\mbhe^*}\right)^{\alpha}+\left(
\frac{\mbhe}{\mbhe^*}\right)^{\beta}} \, .
    \label{eq|Nbhactive}
\end{equation}
At all redshifts, we then determine the parameters
$N_0$, $\alpha$, $\beta$, and $\mbhe^*$ by first computing
the convolution with the input Eddington ratio distribution \PLzM\ and then
finding the best-fit to the AGN luminosity function
\begin{equation}
\Phi(L,z)=\int d \log \lambda P(\lambda|M_{\rm BH},z) \Nacte\, .
\label{eq|PhiLLambdaNact}
\end{equation}

Having \Nact\ at all redshifts we then compute the average accretion
rate from Eq.~\ref{eq|MdotAve} and then compute the black hole mass
function from the continuity equation in Eq.~\ref{eq|conteq}.
The duty cycle at all redshifts and black hole masses
is then simply computed from the ratio of \Nact\ and the total
black hole mass function. The method described above, which we take
as a reference throughout the paper, allows one to describe
the accretion histories of black holes for any continuous input
Eddington ratio distribution. However, it also relies
on assuming an a priori shape for the active mass
function of black holes.

We have developed another method
to solve for the duty cycle
in Eq.~\ref{eq|PhiLLambda} that does not rely on
any a priori shape for the active mass function.
We first assume that the active black holes of mass within $\mbhe$
at a given redshift $z$ with duty cycle $U(\mbhe,z)$ accrete
following a predefined \emph{discrete} Eddington ratio distribution
$P_j(\mbhe,z)$. We take $N$ values of the Eddington ratio
$\lambda=\lambda_j$, with $j=1,..,N$, and set the relative
probability for a black hole to accrete at $\lambda_j$ equal to
$P_j(\mbhe,z)$. The fraction of active black holes accreting at
$\lambda=\lambda_j$ at redshift $z$ is then given by
$P_j(\mbhe,z)U(\mbhe,z)$, and the distribution $P_j$ must be such to satisfy
the condition
\begin{equation}
\sum_j P_j(\mbhe,z)U(\mbhe,z)=U(\mbhe,z)
\label{eq|pjconstraint}
\end{equation}
so that the total fraction of active black holes of mass $\mbhe$
is given by $U(\mbhe,z)$. We then solve
Eq.~(\ref{eq|PhiLLambda}) numerically by computing the duty
cycle at each point $\mbhe$ in the grid as
\begin{eqnarray}
U(\mbhe,z)=\left(P_{j_{MAX}}\Phi_{\rm
BH}(\mbhe,z)\right)^{-1}[\Phi(L,z)|_{L\propto \lambda_{j_{MAX}}\mbhe}-\\
\nonumber
-\sum_{j\neq
j_{MAX}}P_j\Phi_{\rm BH}(\mbhe',z)|_{L\propto \lambda_j \mbhe';\,
\mbhe'>\mbhe}]\, . \label{eq|Uxdiscrete}
\end{eqnarray}
A black hole of mass $\mbhe$ can radiate at the
maximum luminosity $L \propto \lambda_{j_{MAX}}\mbhe$ set by the maximum
assumed Eddington ratio $\lambda_{j_{MAX}}$. More massive black
holes with mass $\mbhe'>\mbhe$ can in principle radiate with similar
luminosity $L$ but at lower Eddington ratios
$\lambda_{j}<\lambda_{j_{MAX}}$, according to the relation $L \propto
\lambda_j \mbhe'$. Therefore Eq.~A4 computes the
duty cycle at a given bin around $\mbhe$ subtracting from the luminosity
function computed at the bin $L \propto \lambda_{j_{MAX}} \mbhe$, the
contributions of more massive black holes with $\mbhe'>\mbhe$ shining within
the same bin around $L$ with lower Eddington ratios.

Note that Eq.~A4 is \emph{implicit}, as in order to
compute $U(\mbhe,z)$ it requires knowledge of the duty cycle at
higher masses $U(\mbhe'>\mbhe,z)$. However, the more luminous sources in the
AGN luminosity function with luminosity $L_{MAX}$ can only be
produced by the more massive black holes in the grid with mass
$\mbhex$ accreting at the maximum Eddington ratio
$\lambda_{j_{MAX}}$. Therefore we can set
\begin{equation}
U(\mbhex,z)=\frac{\Phi(L,z)}{\Phi_{\rm BH}(\mbhex,z)}|_{L\propto
\lambda_{j_{MAX}}\mbhex}\, , \label{eq|Uxdiscretemax}
\end{equation}
and then iteratively solve Eq.~A5 for lower
and lower mass black holes. We have checked that our results
are not dependent on the exact choices for the maximum black hole
mass in the grid $\mbhex$ or the maximum luminosity $L_{MAX}$, as long as they
are sufficiently large (e.g., $\log \mbhex>9.5-10$ and $\log L_{MAX}>48$). For
all our models we set to $\lambda_{j_{MAX}}=1$.

We have checked that the discrete method described above
yields consistent results with the continuous method when the same
input \PLzM\ distribution is adopted.
However, we do not use the discrete method as a reference for several reasons.
Being discrete, it generates oscillations in the predicted black hole
mass function
and duty cycle, absent from the continuous model. Moreover, due to its
discrete
nature, the former technique sometimes yields discontinuities
at late times, especially when a complicated, mass-dependent, mapping
between luminosity and halo mass is assumed.
Other methods to solve Eq.~\ref{eq|conteq} in the presence of broad
Eddington ratio distributions have been discussed by, e.g., Cao \& Li (2008),
although those techniques, based on converting luminosity-based \PL\
to mass-based
ones, work well mainly for Gaussian distributions.

In Section~\ref{subsec|kbolLambda} we discuss models
characterized by a \lam-dependent bolometric
correction. For this class of models
the solution to the continuity equation
is still straightforward even in the presence of broad \PLzM\ distributions.
The same exact procedure detailed above can be extended to
\lam-dependent bolometric
corrections by making use of the
equality $L_{\rm bol}=L_{2-10\, {\rm keV}} K(\lambda)=\lambda L_{\rm Edd}$
and simply assuming a new scale of ``effective'' $\lambda'=
\lambda/K(\lambda)$ that map X-ray luminosities $L_{2-10\, {\rm keV}}$
to Eddington luminosities, and using the X-ray luminosity function
into Eq.~\ref{eq|PhiLLambdaNact} instead of the full bolometric luminosity
function.


\section{The mean relation between black hole and host halo mass}
\label{Appendix|MbhMhaloRelation}

Our models predict, by construction,
the duty cycle and black hole mass function at all times during the evolution.
It is therefore possible to infer the mean relation between black hole
mass and halo
mass via the cumulative relation\footnote{In this work we do not consider any
scatter in the median black hole-halo relation. 
In future work we will probe the most suitable
black hole-halo relation against AGN clustering, scatter, subhalo
accretion histories.
Changing the $M_{\rm BH}-M$ relation would alter the impact
of mergers relative to that discussed in Section~\ref{subsec|Mergers},
but we expect such changes to be small.}
(e.g., Haiman \& Hui 2001; Martini \& Weinberg 2001; Wyithe \& Loeb 2005)
\begin{eqnarray}
\int_{\log \bar{M}}^{\infty} d\log M \Phi_H(M,z)= \nonumber \\
\int_{\log \bar{M}_{\rm BH}}^{\infty}d\log M_{\rm BH} \Phi_{\rm BH}(M_{\rm
BH},z)
\label{eq|cumulRelScatt}
\end{eqnarray}
with both $\Phi_{\rm BH}$
and $\Phi_H$ in units of comoving ${\rm Mpc^{-3}\, dex^{-1}}$ for $H_0=70\,
{\rm km\, s^{-1}\, Mpc^{-1}}$.

The halo mass function $\Phi_H$ is ``corrected'' to include subhaloes as
\begin{equation}
\Phi_H(M,z)=\Phi_{\rm ST}(M,z)+\int n_{\rm SH}(M,M')\Phi_{\rm
ST}(M',z)d\log M'\, , \label{eq|HMFwithSubhalo}
\end{equation}
where $\Phi_{\rm ST}(M,z)$ is the Sheth \& Tormen (1999)\footnote{More recent
and detailed analysis of the halo mass functions have been performed (e.g.,
Tinker et al. 2008); however, we here still use the Sheth \& Tormen recipe
to make contact with previous works and note that the exact choice of
halo mass function does not alter our conclusions.} halo mass function and
$n_{\rm SH}(M,M')$ is the mass function from Giocoli, Tormen \& van den Bosch
(2008) which provides the number of subhaloes with
\emph{unstripped} mass within $M$ and $M+dM$ contained by haloes of
mass $M'$ and $M'+dM'$. Eq.~B2
significantly steepens the halo mass function below $M\sim 10^{12}\,
{\rm h^{-1}\, M_{\odot}}$, as it increases the number of lower mass
haloes more, thus somewhat altering the mean correlation between
haloes and black holes in this mass regime.

\section{Computing black hole mergers}
\label{Appendix|Mergers}

In the presence of mergers
two additional terms should be added on the right hand side
of Eq.~\ref{eq|conteq}
\begin{eqnarray}
S_{\rm in}=\frac{1}{4}\, \int_{\xi_{\rm min}}^{1} d\xi \left( \frac{P_{\rm
merg,z}}{\Delta t}(\xi, M)n_h \left[M \left(M_{\rm BH}', z \right),z
\right]\frac{dM}{dM_{\rm BH}} \right) \\ +\nonumber
d\xi \left(\frac{P_{\rm merg,z}}{\Delta t}(\xi, M)n_h \left[M \left(M_{\rm
BH}'', z \right),z \right]\frac{dM}{dM_{\rm BH}}  \right)
\label{eqApp|Sin}
\end{eqnarray}
is the merger rate of incoming smaller mass black holes with
mass $M_{\rm BH}'=M_{\rm BH}\xi/(1+\xi)$ and $M_{\rm BH}''=M_{\rm BH}/(1+\xi)$
that
merge into a black hole of final mass $M_{\rm BH}$, and
\begin{eqnarray}
S_{\rm out}=\frac{1}{2}\, \int_{\xi_{\rm min}}^{1} d\xi \left( \frac{P_{\rm
merg,z}}{\Delta t}(\xi, M) n_h \left [M \left(M_{\rm BH}', z \right),z
\right]\frac{dM}{dM_{\rm BH}} \right) \\ \nonumber +d\xi \left(\frac{P_{\rm
merg,z}}{\Delta t}(\xi, M)n_h \left[M \left(M_{\rm BH}'', z \right),z
\right]\frac{dM}{dM_{\rm BH}}  \right)
\label{eqApp|Sout}
\end{eqnarray}
is the merger rate of black holes with initial mass $M_{\rm BH}$ that merge
into more massive black holes of mass
$M_{\rm BH}'=M_{\rm BH}(1+\xi)/\xi$ and $M_{\rm BH}''=M_{\rm BH}(1+\xi)$.
In both Eqs.~C1 and C2 we set $\xi_{\rm min}=0.1$ or $\xi_{\rm min}=0.3$,
and add the factor of $1/2$ to avoid double counting; an additional factor of 1/2 is present
in $S_{\rm in}$ to take into account that two black holes merge into one.

The probabilities of halo mergers per unit time as a function of mass ratio, redshift,
and remnant mass are taken from Fakhouri \& Ma (2008).
Following Shen (2009), we then insert a delay between host dark halo and
host galaxy merger rate determined by the dynamical
friction time of typical haloes of that mass. The probability of black hole
mergers per unit time is then simply given by the
``delayed'' halo merger rate, i.e.,
\begin{equation}
\frac{P_{\rm merg,z'>z}}{\Delta t}(\xi, M)n_h [M (M_{\rm BH},
z),z]=B_h[M,\xi(z),z]\, ,
\label{eqApp|ratemergers}
\end{equation}
where $z'$ is the actual epoch of the merger of the progenitor haloes
(see Shen 2009 for details).
By simply knowing, at each timestep, the mapping between infalling halo
mass and its central black hole (given by Eq.~[\ref{eq|cumulRelScatt}]),
we can then compute the expected average rate
for any black hole merger event.

\section{A comprehensive collection of active galaxies at different
masses and redshifts}
\label{Appendix|AGNdutyCycles}


In this Appendix we present an overview of a detailed literature search
aimed at extracting an approximate estimate of the duty cycle
of active galaxies at redshifts $z\gtrsim 1$ above some given luminosity.

At $z\sim 2$ several estimates of the number density
of certain type of galaxies, along with their AGN fraction, have
been measured. Caputi et al. (2006) estimated the cumulative number
density of $K$-selected galaxies at high redshifts from the
GOODS-Chandra Deep Fields to be $\lesssim 2\times 10^{-4}\, {\rm
Mpc^{-3}}$. They also
found that the majority of the most massive galaxies in their sample,
with stellar masses $> 2.5\times 10^{11}\, M_{\odot}$, are
ultraluminous infrared galaxies, out of which at least $15$\% show
AGN signatures from their X-ray luminosities. They claim the AGN
fraction to decrease by about a factor of two moving to lower masses
and lower infrared luminosities, although this may be partly due to
selection effects (Caputi et al. 2006).

In order to extract an absolute value of the AGN fraction from the
AGN percentage revealed in the Caputi et al. (2006) sample, we need
to correct for the fact that the density of galaxies in their sample
may be below the actual total number of galaxies of the same mass at
the same redshift. Drory et al. (2005) derived the
total galaxy stellar mass function at $2<z<2.5$ from \emph{I} and
\emph{K} selected galaxies in the FORS Deep and GOODS Fields,
respectively (see also Fontana et al. 2006).
More recently, near-IR estimates come from
P\'{e}rez-Gonz\'{a}lez et al. (2008) and Marchesini et al. (2009).
Given the large clustering of high redshift massive galaxies (e.g., Foucaud et
al. 2007; Magliocchetti et al. 2007; Quadri et al. 2007)
field-to-field variations may significantly affect these results
(e.g., Marchesini et al. 2007), but the nice agreement with the
estimate by van Dokkum et al. (2006) derived from a large area
is also reassuring.

We thus find that the $K$-selected galaxy number density derived by
Caputi et al. (2006) amounts to about 60\% of the cumulative number
density of galaxies in the stellar mass function with mass above
$10^{11}\, M_{\odot}$. We therefore take a mean of $0.15\times
0.6\sim 0.1$ as representative of the AGN fraction in galaxies with
at $z \sim 2$ of this mass.

Erb et al. (2006) find that only 5 out of 114 Lyman galaxies host
AGNs, consistent with previous claims from similar samples by, e.g.,
Nandra et al. (2002, 2005), Gawiser et al. (2007). Adelberger \&
Steidel (2005) estimated an AGN duty cycle of about $\sim 15\%$
(although with large error bars) by combining clustering
measurements and space densities of AGNs within their large sample
of star forming galaxies at $1.8\lesssim z \lesssim 3.5$. Kriek et
al. (2007) instead found from a sample of 20 $K$-selected galaxies,
that the fraction of AGNs inferred from several emission line
diagnostics, was $\sim 20\%$ up to 40\% for the more massive
galaxies, and rapidly dropping at lower masses. 
The Kriek et al. results are
at variance with the AGN fraction estimated in UV-selected galaxies
by Erb et al. (2006) but in reasonable agreement with Caputi et al.
(2006). However, Kriek et al. (2007) proved that their estimates are
actually consistent with those from UV sample. By binning the Erb et
al. sample in stellar mass they found in fact that the AGN fraction
in galaxies with stellar mass of $10^{11}\, M_{\odot}$ is about
$16\%$ rising to $\sim 40\%$ at higher masses. 
Similar results were also claimed by Reddy et al. (2005), shown
by a filled circle in the same Figure, who estimated that $\sim 25\%$
of their $K_s$-selected galaxies in the deep \emph{Chandra} fields
are active. Note that the Kriek et al. and Erb et al. results
are made somewhat uncertain by the fact that we do not have a proper
estimate of the comoving abundance of galaxies in their samples.
Papovich et al. (2005) also estimated an AGN fraction of $25\%$
among a sample of distant red galaxies, which reduces to an overall
$20\%$ if distant red galaxies comprise about 70\% of the overall
population of massive high redshift galaxies (van Dokkum et al.
2006; Marchesini et al. 2007).

The majority ($\gtrsim 75\%$) of the Submillimeter galaxies observed
with deep SCUBA surveys at $850\mu$ and ultradeep X-ray observations
from the 2Msec Chandra Deep Field North, have been observed to host
a central AGN (Alexander et al. 2005). The mean BH mass has been
calibrated to be $\sim 10^8\, M_{\odot}$ and the mean Eddington
ratio distribution peaked around $0.2<\lambda<1$ (Borys et al. 2005;
Alexander et al. 2008). However, assuming the comoving density of SCUBA
galaxies to be less than 1/4 the cumulative number density of
$K$-selected massive galaxies (e.g., Chapman et al. 2005),
we get an average AGN fraction of $\sim 17\%$, comparable to the result by
Caputi et al. (2006).

At lower redshifts estimates of the AGN fraction in large samples of
galaxies have been recently estimated by Bundy et al. (2008), from
DEEP2 and AEGIS surveys. In their sample AGNs have been detected in
the 2-10 keV hard band of Chandra and therefore they are
representative of the overall AGN population (except for highly
obscured, Compton thick AGNs). We find the Bundy
et al. (2008) stellar mass function to overall
in good agreement with other studies, within a factor of $\sim 2$.
The ratio between the active and inactive stellar mass
functions estimated by Bundy et al. (2008), yields a duty cycle
increasing with black hole mass. The AGN fraction is $\sim
10\%$ raising to $\sim 50\%$ at $z\sim 1.1$, and from 0.01 to 0.1 at
$z\sim 0.5$. This result is in good agreement with the estimates by
Lehmer et al. (2007) who find that about 5\% of the early-type
galaxies in the Extended Chandra Deep Field-South contain AGNs in
the local universe, increasing as $(1+z)^3$ at higher redshifts
(i.e., about 17\% at $z\sim 0.5$). A closer inspection by Silverman
et al. (2008b) on an X-ray selected sample at redshifts $0.63\lesssim
z \lesssim 0.76$, has confirmed an AGN fraction of (15$\pm$ 5)\%,
decreasing to $\sim 2-6\%$ for lower luminosity galaxies. Similarly,
Shi et al. (2008) find a lower limit of $2\%$ for X-ray selected
AGNs at $z\sim 0.5$ in low mass galaxies increasing to $\sim 10\%$
in the more massive systems, in good agreement with what
inferred by Bundy et al. (2008). The Bundy et al. (2008) average
estimates also agree well with the Montero-Dorta et al. (2009)
results on the AGN fraction of galaxies residing in different
environments at $z\sim 1$ from AEGIS and DEEP2.


\begin{thebibliography}{}
\bibitem{} Adelberger K. L., Steidel C. C., 2005, ApJ, 630, 50
\bibitem{} Adelman-McCarthy J.~K. et al., 2006, ApJS, 162, 38
\bibitem{} Aird J. et al., 2011, arXiv:1107.4368
\bibitem{} Aird J. et al., 2010, MNRAS, 401, 2531
\bibitem{} Alexander D. M. et al., 2005, ApJ, 632, 736
\bibitem{} Alexander D. M. et al., 2008, AJ, 135, 1968
\bibitem{} Alexander D.~M. et al., 2011, arXiv:1106.1443
\bibitem{} Babi\'{c} A., Miller L., Jarvis M. J., Turner T. J., Alexander
D. M., Croom S. M., 2007, A\&A, 474, 755
\bibitem{} Ballo L. et al., 2007, ApJ, 667, 97
\bibitem{} Best P.~N., Kauffmann G., Heckman T.~M. et al., 2005, MNRAS, 362, 25
\bibitem{} Bonoli S., Marulli F., Springel V., White S.~D.~M., Branchini E.,
Moscardini L.,\ 2009, MNRAS, 606
\bibitem{} Bonoli S., Shankar F., White S.~D.~M., Springel V., Wyithe
J.~S.~B., 2010, MNRAS, 404, 399
\bibitem{} Borys C., Smail I., Chapman S. C.,
Blain A. W., Alexander D. M., Ivison R. J., 2005, ApJ, 635, 853
\bibitem{} Bundy K. et al., 2008, ApJ, 681, 931
\bibitem{} Cao X., Li F., 2008, MNRAS, 390, 561
\bibitem{} Cao X., 2010, ApJ, 725, 388
\bibitem{} Capetti A., 2011, arXiv:1109.6196
\bibitem{} Caputi K. I. et al., 2006, A\&A, 454, 143
\bibitem{} Cavaliere A., Morrison P., Wood K., 1971, ApJ, 170, 223
\bibitem{} Cavaliere A., Vittorini V., 2000, ApJ, 543, 599
\bibitem{} Chapman S.~C., Blain A.~W., Smail I., Ivison R.~J., 2005, ApJ,
622, 772
\bibitem{} Conroy C., \& White M., 2012, arXiv:1208.3198
\bibitem{} Croom S. M. et al., 2005, MNRAS, 356, 415
\bibitem{} Croom S.~M. et al., 2009, MNRAS, 399, 1755
\bibitem{} Dai X., Chartas G., Eracleous M., Garmire G. P., 2004, ApJ, 605, 45
\bibitem{} Davis S.~W., \& Laor A., 2011, ApJ, 728, 98
\bibitem{} Di Matteo T., Springel V., Hernquist L., 2005 Nat,
433, 604
\bibitem{} Draper A.~R. \& Ballantyne D.~R., 2010, ApJL, 715, L99
\bibitem{} Drory N., Salvato M., Gabasch A., Bender R., Hopp U., Feulner
G., Pannella M., 2005, ApJ, 619, 131
\bibitem{} Eddington A. S., 1922, MNRAS, 83, 32
\bibitem{} Elvis M. et al., 1994, ApJS, 95, 1
\bibitem{} Erb D. K., Shapley A. E., Pettini M., Steidel C. C., Reddy N. A.,
Adelberger K. L., 2006, ApJ, 644, 813
\bibitem{} Fakhouri O., Ma C.-P.,\ 2008, MNRAS, 386, 577
\bibitem{} Fanidakis N., Baugh C.~M., Benson A.~J., Bower R.~G., Cole S.,
Done C., Frenk C.~S., 2011, MNRAS, 410, 53
\bibitem{} Ferrarese L., Merritt D., 2000, ApJ, 539, L9
\bibitem{} Ferrarese L., Ford H., 2005, SSRv, 116, 523
\bibitem{} Fine S. et al., 2008, MNRAS, 390, 1413
\bibitem{} Fiore F. et al., 2011, arXiv:1109.2888
\bibitem{} Fontana A. et al., 2006, A\&A, 459, 745
\bibitem{} Foucaud S. et al., 2007, MNRAS, 376, 20
\bibitem{} Gawiser E. et al., 2007, ApJ, 671, 278
\bibitem{} Gebhardt K. et al., 2000, ApJL, 539, 13
\bibitem{} Giocoli C., Tormen G., van den Bosch F.~C., 2008, MNRAS, 386, 2135
\bibitem{} Gonz{\'a}lez V., Labb{\'e} I., Bouwens R.~J., Illingworth G.,
Franx M., Kriek M., Brammer G.~B., 2010, ApJ, 713, 115
\bibitem{} Goulding A.~D., Alexander D.~M., Lehmer B.~D., Mullaney J.~R.,
2010, MNRAS, 662
\bibitem{} Graham A. W., Driver S. P., Allen P. D., Liske
J., 2007, MNRAS, 378, 198
\bibitem{} Granato G. L., Silva L., Lapi A., Shankar F., De Zotti G., Danese
L., 2006, MNRAS, 368L, 72
\bibitem{} Greene J.~E., Ho L.~C., 2007, ApJ, 667, 131
\bibitem{} Greene J.~E., Ho L.~C., 2009, ApJ, 704, 1743
\bibitem{} Grier C.~J., Mathur S.,
Ghosh H., Ferrarese L., 2011, ApJ, 731, 60
\bibitem{} Haiman Z., Hui L., 2001, ApJ, 547, 27
\bibitem{} H\"{a}ring N., Rix H. W.,  2004, ApJ, 604, 89
\bibitem{} Heckman T. M. et al., 2004, ApJ, 613, 109
\bibitem{} Heckman T.~M., Ptak A., Hornschemeier A., Kauffmann G., 2005,
ApJ, 634, 161
\bibitem{} Hickox R.~C. et al., 2009, ApJ, 696, 891
\bibitem{} Ho L.~C., 2004, Multiwavelength AGN Surveys, 153
\bibitem{} Hopkins A., Beacom J. F., 2006, ApJ, 651, 142
\bibitem{} Hopkins P. F., Hernquist L., Cox T. J., Robertson B., Di Matteo
T., Springel V., 2006, ApJ, 639, 700
\bibitem{} Hopkins P. F., Richards G. T., Hernquist L.,
2007, ApJ, 654, 731
\bibitem{} Hopkins P.~F., Hernquist L., 2009, ApJ, 698, 1550
\bibitem{} Hughes S. A., Blandford R. D., 2003, ApJ, 585, L101
\bibitem{} Islam R.~R., Taylor J.~E., Silk J., 2004, MNRAS, 354, 427
\bibitem{} Karim A. et al., 2011, ApJ, 730, 61
\bibitem{} Kauffmann G., Haehnelt M., 2000, MNRAS, 311, 576
\bibitem{} Kauffmann G. et al., 2003, MNRAS, 346, 1055
\bibitem{} Kauffmann G., Heckman T.~M., 2009, MNRAS, 397, 135
\bibitem{} Kelly B.~C., Vestergaard M., Fan X., Hopkins P., Hernquist L.,
Siemiginowska A., 2010, ApJ, 719, 1315
\bibitem{} Kennicutt R.~C. et al., 2003, PASP, 115, 928
\bibitem{} Kisaka S., Kojima Y., 2010, MNRAS, 405, 1285
\bibitem{} Kocsis B., Sesana A., 2011, MNRAS, 411, 1467
\bibitem{} Kollmeier J. A. et al., 2006, ApJ, 648, 128 (K06)
\bibitem{} Kochanek, C.S., Eisenstein, D. J., Cool, R. J., et al.\ 2011,
           ApJS, submitted, arXiv:1110.4371
\bibitem{} Kriek M. et al., 2007, ApJ, 669, 776
\bibitem{} Kulkarni G., Loeb A., 2011, arXiv:1107.0517
\bibitem{} Lamastra A., Matt G., Perola G.~C., 2006, A\&A, 460, 487
\bibitem{} Laor A., Davis S., 2011, arXiv:1110.0653
\bibitem{} Lapi A., Shankar F., Mao J., Granato G. L.,
Silva L., De Zotti G., Danese L., 2006, ApJ, 650, 42
\bibitem{} Lauer T. R., Tremaine S., Richstone D., Faber S.
M., 2007, ApJ, 670, 249
\bibitem{} Lehmer B.~D. et al., 2007, ApJ, 657, 681
\bibitem{} Lusso E. et al., 2010, A\&A, 512, A34
\bibitem{} Metcalf R.~B., Magliocchetti M., 2006, MNRAS, 365, 101
\bibitem{} Magliocchetti M., Silva L.,
Lapi A., de Zotti G., Granato G. L., Fadda D., Danese L.,
2007, MNRAS, 375, 1121
\bibitem{} Magorrian J. et al., 1998, AJ, 115, 2285
\bibitem{} Mainieri V. et al., 2011, arXiv:1105.5395
\bibitem{} Marchesini D. et al., 2007, ApJ, 656, 42
\bibitem{} Marchesini D., van Dokkum P.~G., F{\"o}rster Schreiber N.~M.,
Franx M., Labb{\'e} I., Wuyts S., 2009, ApJ, 701, 1765
\bibitem{} Marconi A., Risaliti G., Gilli R., Hunt L. K.,
Maiolino R., Salvati M., 2004, MNRAS, 351, 169
\bibitem{} Marconi A. et al., 2008, ApJ, 678, 693
\bibitem{} Martini P., Weinberg D. H., 2001, ApJ, 547, 12
\bibitem{} Marulli F., Bonoli S., Branchini E., Moscardini L., Springel
V., 2008, MNRAS, 385, 1846
\bibitem{} Matteucci F., 1994, A\&A, 288, 57
\bibitem{} McLure R. J., Dunlop. J. S., 2004, MNRAS, 352, 1390
\bibitem{} McLure R.~J., Jarvis M.~J., 2004, MNRAS, 353, L45
\bibitem{} Merloni A., 2004, MNRAS, 353, 1035
\bibitem{} Merloni A., Rudnick G., Di Matteo T., 2004, MNRAS, 354, L37
\bibitem{} Merloni A., Heinz S., 2008, MNRAS, 388, 1011
\bibitem{} Metcalf R.~B., Magliocchetti M., 2006, MNRAS, 365, 101
\bibitem{} Montero-Dorta A.~D. et al., 2009, MNRAS, 392, 125
\bibitem{} Nandra K., Mushotzky R. F., Arnaud K., Steidel C. C.,
Adelberger K. L., Gardner J. P., Teplitz H. I., Windhorst R.
A., 2002, ApJ, 576, 625
\bibitem{} Nandra K., Laird E. S., Steidel C. C., 2005, MNRAS,
\bibitem{} Netzer H., Trakhtenbrot B., 2007, ApJ, 654, 754
\bibitem{} Netzer H., Lira P., Trakhtenbrot B., Shemmer O., Cury I.,
2007, ApJ,
671, 1256
\bibitem{} Netzer H., 2009, ApJ, 695, 793
\bibitem{} Noeske K.~G., 2009, Astronomical Society of the Pacific Conference
Series, 419, 298
\bibitem{} Osmer P.~S., 1982, ApJ, 253, 28
\bibitem{} Papovich C., Dickinson M., Giavalisco M.,
Conselice C. J., Ferguson H. C., 2005, ApJ, 631, 101
\bibitem{} P{\'e}rez-Gonz{\'a}lez P.~G. et al., 2008, ApJ, 675, 234
\bibitem{} Quadri R. et al., 2007, ApJ, 654, 138
\bibitem{} Rafiee A., Hall P.~B., 2011, ApJS, 194, 42
\bibitem{} Raimundo S.~I., Fabian A.~C., 2009, MNRAS, 396, 1217
\bibitem{} Raimundo S.~I., Fabian A.~C., Vasudevan R.~V., Gandhi P., Wu J., 2011, arXiv:1109.6225
\bibitem{} Reddy N. A., Erb D. K., Steidel C. C., Shapley A. E.,
Adelberger K. L., Pettini M., 2005, ApJ, 633, 748
\bibitem{} Richards G.~T. et al., 2006, ApJS, 166, 470
\bibitem{} Ross N.~P. et al., 2009, ApJ, 697, 1634
\bibitem{} Rovilos E., Georgantopoulos I., 2007, A\&A, 475, 115
\bibitem{} Saez C. et al., 2008, AJ, 135, 1505
\bibitem{} Salpeter E. E., 1964, ApJ, 140, 796
\bibitem{} Salucci P., Szuszkiewicz E., Monaco P., Danese L.,
1999, MNRAS, 307, 637
\bibitem{} Scannapieco E., Oh S. P., 2004, ApJ, 608, 62
Joo S. J., Yi S. K., Silk
\bibitem{} Schulze A., Wisotzki L., 2010, A\&A, 516, 87
\bibitem{} Schulze A., Gebhardt K., 2011, ApJ, 729, 21
\bibitem{} Shankar F., Salucci P., Granato G. L., De Zotti G., Danese
L., 2004, MNRAS, 354, 1020
\bibitem{} Shankar F., Cavaliere A., Cirasuolo M., Maraschi L., 2008a, ApJ,
676, 131
\bibitem{} Shankar F., Dai X., Sivakoff G. R., 2008b, ApJ, 687, 859
\bibitem{} Shankar F., Weinberg D. H., Miralda-Escud\'{e}, J.
2009, ApJ, 690, 20 (SWM)
\bibitem{} Shankar F., 2009, New Astronomy Reviews, 53, 57
\bibitem{} Shankar F., Bernardi M., Haiman Z., 2009, ApJ, 694, 867
\bibitem{} Shankar F., Marulli F., Bernardi M., Dai X., Hyde J.~B., Sheth
R.~K., 2010a, MNRAS, 403, 117
\bibitem{} Shankar F., Crocce M., Miralda-Escud\'{e} J., Fosalba
P., Weinberg D. H., 2010b, ApJ, 718, 231
\bibitem{} Shankar F., Weinberg D. H., Shen Y., 2010c, MNRAS, 406, 1959
\bibitem{} Shankar F., Marulli F., Bernardi M., Boylan-Kolchin M., Dai X.,
Khochfar S., 2010d, MNRAS, 405, 948
\bibitem{} Shankar F., Sivakoff G.~R., Vestergaard M., Dai X., 2010e, MNRAS,
401, 1869
\bibitem{} Shankar F., 2010, IAU Symposium, 267, 248
\bibitem{} Shankar F., Marulli F., Bernardi M., Mei S., Meert A., Vikram V.,
2011, arXiv:1105.6043
\bibitem{} Shen Y. et al., 2007, AJ, 133, 2222
\bibitem{} Shen Y., Greene J. E.,
Strauss M. A., Richards G. T., Schneider D. P., 2008, ApJ, 680,
169
\bibitem{} Shen Y., 2009, ApJ, 704, 89
\bibitem{} Shen Y. et al., 2009, ApJ, 697, 1656
\bibitem{} Shen Y., Kelly B.~C., 2010, ApJ, 713, 41
\bibitem{} Shen Y., Kelly B.~C., 2011, arXiv:1107.4372
\bibitem{} Sheth R.~K., Tormen G., 1999, MNRAS, 308, 119
\bibitem{} Shi Y., Rieke G., Donley J., Cooper M., Willmer C., Kirby E.,
2008, ApJ, 688, 794
\bibitem{} Silverman J.~D. et al., 2008a, ApJ, 679, 118
\bibitem{} Silverman J. D. et al., 2008b, ApJ, 675, 1025
\bibitem{} Silverman J. D. et al., 2009, ApJ, 696, 396
\bibitem{} Small T. A., Blandford R. D., 1992, MNRAS, 259, 725
\bibitem{} So\l tan A., 1982, MNRAS, 200, 115
\bibitem{} Smith R. E. et al., 2003, MNRAS, 341, 1311
\bibitem{} Steed A., Weinberg D. H., 2003, astroph/0311312
\bibitem{} Steinhardt C.~L., Elvis M., 2010a, MNRAS, 402, 2637
\bibitem{} Steinhardt C.~L., Elvis M., 2010b, MNRAS, L72
\bibitem{} Tamura N., Ohta K., Ueda Y., 2006, MNRAS, 365, 134
\bibitem{} Tinker J., Kravtsov A.~V., Klypin A., Abazajian K., Warren M.,
Yepes G., Gottl{\"o}ber S., Holz D.~E., 2008, ApJ, 688, 709
\bibitem{} Trakhtenbrot B., Netzer H., Lira P., Shemmer O., 2011, ApJ, 730, 7
\bibitem{} Trump J.~R. et al., 2011, ApJ, 733, 60
\bibitem{} Ueda Y., Akiyama M., Ohta K., Miyaji T., 2003, ApJ, 598, 886
\bibitem{} van Dokkum P. G. et al., 2006, ApJ, 638, 59
\bibitem{} Vasudevan R. V., Fabian A. C., 2007, MNRAS, 381, 1235
\bibitem{} Vasudevan R.~V., Fabian A.~C., 2009, MNRAS, 392, 1124
\bibitem{} Vestergaard M., 2002, ApJ, 571, 733
\bibitem{} Vestergaard M., 2004, ApJ, 601, 676
\bibitem{} Vika M., Driver S.~P., Graham A.~W., Liske J., 2009, MNRAS,
400, 1451
\bibitem{} Vittorini V., Shankar F., Cavaliere A., 2005, MNRAS,
363, 1376
\bibitem{} Volonteri M., Madau P., Quataert E., Rees M.~J., 2005, ApJ,
620, 69
\bibitem{} Volonteri M., Sikora M., Lasota J. P., 2007, ApJ,
667, 704
\bibitem{} Wandel A., Peterson B.~M., Malkan M.~A., 1999, ApJ, 526, 579
\bibitem{} Wang J.-M. et al., 2009, ApJL, 697, L141
\bibitem{} Weinmann S.~M., Neistein E., Dekel A., 2011, arXiv:1103.3011
\bibitem{} White M., Martini P., Cohn J.~D., 2008, MNRAS, 390, 1179
\bibitem{} Willott C.~J. et al., 2010a, AJ, 139, 906
\bibitem{} Willott C.~J. et al., 2010b, arXiv:1006.1342
\bibitem{} Wyithe J. S. B., Loeb A., 2003, ApJ, 595, 614
\bibitem{} Wyithe J. S. B., Loeb A., 2005, ApJ, 621, 95
\bibitem{} Xue Y.~Q. et al., 2010, ApJ, 720, 368
\bibitem{} Yoo J., Miralda-Escud{\'e} J., 2004, ApJL, 614, L25
\bibitem{} Yoo J., Miralda-Escud{\'e} J., Weinberg D.~H., Zheng Z., Morgan
C.~W., 2007, ApJ, 667, 813
\bibitem{} Yu Q., Tremaine S., 2002, MNRAS, 335, 965
\bibitem{} Yu Q., Lu Y., 2004, ApJ, 602, 603
\bibitem{} Yu Q., Lu Y., 2008, ApJ, 689, 732
\bibitem{} Zheng X.~Z. et al., 2009, ApJ, 707, 1566
\end{thebibliography}
\end{document}